\documentclass[12pt,a4paper]{article}

\usepackage{amsmath}
\usepackage{amssymb}
\usepackage{graphicx}
\usepackage{dsfont}
\usepackage{overpic}
\usepackage{cite}
\usepackage{tabls}

\let\oldappendix=\appendix
\let\oldsection=\section
\renewcommand{\appendix}{\oldappendix%
\def\theequation{\Alph{section}.\arabic{equation}}%
\renewcommand{\section}{\setcounter{equation}{0}\oldsection}}

\newcommand{\beq}{\begin{equation}}
\newcommand{\eeq}{\end{equation}}
\newcommand{\beqa}{\begin{eqnarray}}
\newcommand{\eeqa}{\end{eqnarray}}
\newcommand{\no}{\nonumber}

\newcommand{\tr}{\mbox{tr}}

\newcommand{\newop}[2]{\def#1{\mathop{\mathrm{#2}}\nolimits}}
\newop{\artanh}{artanh}
\newop{\det}{det}
\newop{\tr}{tr}
\newop{\diag}{diag}
\newop{\Re}{Re}
\newop{\Im}{Im}

\newcommand{\MeV}{\,\mathrm{MeV}}
\newcommand{\GeV}{\,\mathrm{GeV}}
\newcommand{\fm}{\,\mathrm{fm}}
\newcommand{\Lagr}{\mathcal{L}}

\newcommand{\vs}{\vspace{-0.25cm}}

\begin{document}

\hfill {\tiny HISKP-TH-06/15, FZJ-IKP(TH)-2006-16}

\hfill 

\bigskip\bigskip

\begin{center}

{{\Large\bf $\mbox{\boldmath$K^- p$}$ scattering length
            from scattering experiments}}

\end{center}

\vspace{.4in}

\begin{center}
{\large B.~Borasoy\footnote{email: borasoy@itkp.uni-bonn.de}$^{a}$,
        U.-G.~Mei{\ss}ner\footnote{email: meissner@itkp.uni-bonn.de}$^{a, b}$,
        R.~Ni{\ss}ler\footnote{email: rnissler@itkp.uni-bonn.de}$^{a}$}

\bigskip

\bigskip

$^{a}$Helmholtz-Institut f\"ur Strahlen- und Kernphysik (Theorie), \\
Universit\"at Bonn,
Nu{\ss}allee 14-16, D-53115 Bonn, Germany \\[0.4cm]

$^b$Institut f\"ur Kernphysik (Theorie), Forschungszentrum J\"ulich, \\
D-52425 J\"ulich, Germany \\[2ex]

\vspace{.2in}

\end{center}

\vspace{.7in}

\thispagestyle{empty} 

\begin{abstract}
The strong $K^- p$ scattering length is extracted
within chiral SU(3) unitary approaches from a very large variety of
fits to low-energy $K^- p$ scattering data.
Very good overall agreement with available scattering data is obtained and the resulting
scattering length is compared with the new accurate kaonic hydrogen data
from DEAR. The pole structures of the obtained fits to experiment are critically examined.
\end{abstract}\bigskip



\vfill

\section{Introduction}\label{sec:Intro}

The approximate chiral symmetry of the light quarks $u,d,s$ in quantum chromo dynamics
plays a crucial role in describing low-energy hadronic processes. Chiral
SU(3) symmetry is exact in the limit of vanishing light quark masses, but finite
quark masses, $m_u, m_d, m_s$, induce explicit symmetry-breaking.
While for the very light quarks $u,d$ the symmetry-breaking corrections are
in general small, it remains unclear whether the strange quark is still light enough
to be in the chiral regime.

In this respect, the $\bar{K}N$ system provides a good testing ground for chiral SU(3) dynamics
and the role of explicit chiral symmetry-breaking due to the strange quark mass.
High-precision $K^-p$ threshold data, such as $K^-p$ scattering, the $\pi \Sigma$
mass spectrum and the precisely measured $K^-p$ threshold decay ratios
set important constraints for theoretical approaches. Recently they have  
been supplemented by the new accurate results for the strong interaction shift and width
of kaonic hydrogen from the DEAR experiment \cite{DEAR} which reduced both the mean values
and error ranges of the previous KEK experiment \cite{KEK}.

The existence of the $\Lambda(1405)$ resonance in the $K^-p$ channel just
below its threshold makes the loop expansion of chiral perturbation theory
(ChPT) inapplicable.
In this regard, the combination of ChPT with non-perturbative coupled-channel 
techniques based on driving terms of the chiral SU(3) 
effective Lagrangian has proven useful 
by generating the $\Lambda(1405)$ dynamically as an I = 0  $\bar{K}N$ quasibound state 
and a resonance in the $\pi\Sigma$ channel.

However, the recent and precise DEAR measurement appears to be in disagreement
with the $K^-p$ scattering length derived from scattering data as pointed out
in \cite{MRR}. A thorough investigation of the low-energy $K^-p$ interaction
within chiral unitary approaches has reinforced the question of consistency of the
DEAR experiment with $K^-p$ scattering data \cite{BNW1}.
In contrast, it was claimed very recently in \cite{OPV, Oller} that within a 
chiral unitary approach both the scattering and the DEAR data can be accommodated.

It is evident from these recent investigations that the $K^-p$ system
remains a topic of great interest and is under lively discussion, see also \cite{BNW2, OPV2}.
The aim of the present work is to shed some more light on this issue by
providing a conservative range for the $K^-p$ scattering length 
constrained solely from $K^-p$ scattering data. This is accomplished within different
variants of chiral unitary approaches and by performing a very large number of fits to
experiment, in order to reduce the inherent model-dependence
of these frameworks. The obtained realistic range for the $K^-p$ scattering length 
is then compared with the one derived from the DEAR and KEK experiments.

This work is organized as follows. In the following section we present the basic
formalism. Our results for $K^-p$ scattering are shown in Section~\ref{sec:res}. 
Section~\ref{sec:poles} contains an overview of the
pole structures of the obtained solutions in the $\Lambda(1405)$ region.
We summarize our findings in Sec.~\ref{sec:summary}. Some of the detailed
results of this investigation are collected and displayed in the appendix.

\section{Formalism}\label{sec:Form}

In this section we illustrate the underlying features of the approach.
Details of the formalism have already been presented in previous works \cite{BNW1}
and will not be repeated here.
The starting point is the chiral effective
Lagrangian ${\cal L} ={\cal L}_\phi + {\cal L}_{\phi B}$ 
which describes the coupling of
the pseudoscalar meson octet $(\pi,K,\eta)$ to the ground state
baryon octet $(N,\Lambda, \Sigma, \Xi)$.
Both the purely mesonic piece ${\cal L}_\phi$ and the meson-baryon Lagrangian ${\cal L}_{\phi B}$ 
are employed up to second chiral order.

Due to the nearby $\Lambda(1405)$ 
resonance unitarity effects from final state interactions are important
for $\bar{K} N$ scattering and must be included in a non-perturbative fashion.
To this aim, the relativistic effective Lagrangian is utilized to compute the tree
level amplitude $V_{j b, i a}(s, \Omega; \sigma, \sigma')$ of the meson-baryon 
scattering processes $\phi_i B_{a}^{\sigma} \to  \phi_j B_{b}^{\sigma'}$ (with spin 
indices $\sigma$, $\sigma'$) at invariant energy squared $s$. 
This amplitude is the driving term in the coupled-channels 
integral equation determining the meson-baryon $T$-matrix.

\begin{figure}
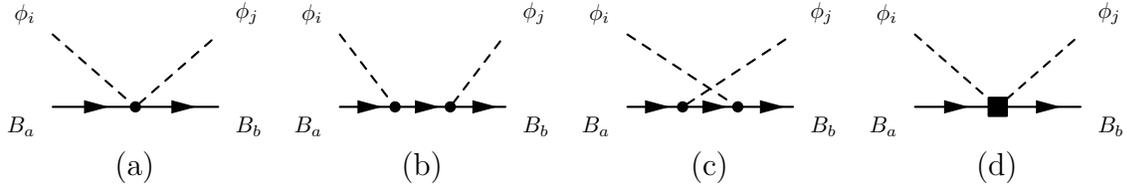

\centering
\begin{tabular}{cccc}
\includegraphics[width=0.2\textwidth]{feynps.1} & 
\includegraphics[width=0.2\textwidth]{feynps.3} &  
\includegraphics[width=0.2\textwidth]{feynps.4} &
\includegraphics[width=0.2\textwidth]{feynps.2} \\
(a) & (b) & (c) & (d)
\end{tabular}
\caption{Shown are the leading order contact interaction (a), the direct (b) and crossed (c)
         Born term as well as the next-to-leading order contact interaction (d).
         Solid and dashed lines represent baryons and pseudoscalar mesons, respectively.}
\label{fig:feyns}
\end{figure}

In the literature, the effective meson-baryon Lagrangian has been
used as interaction kernel at different levels of sophistication. 
While only the Weinberg-Tomozawa term
from the covariant derivative is taken, e.g., in \cite{OR}, the 
Born terms are included in \cite{OM}. In \cite{KSW2}, on the other hand, the Lagrangian
of second chiral order is added which yields additional contact interactions.
In order to provide an estimate of the model-dependence of such approaches,
we will discuss the following three different choices for the amplitude 
$V_{j b, i a}(s, \Omega; \sigma, \sigma')$.
First, only the leading order contact (Weinberg-Tomozawa) term is taken 
into account, see Figure~\ref{fig:feyns}a. 
Subsequently, the Born diagrams are included, see Figs.~\ref{fig:feyns}b and c.
In the third approach we add the contact interactions from the Lagrangian of second 
chiral order, $\Lagr_{\phi B}^{(2)}$ (Fig.~\ref{fig:feyns}d).
For brevity, we will refer to these variants as ``WT'' (Weinberg-Tomozawa),
``WTB'' (Weinberg-Tomozawa + Born diagrams) and ``full'' (including also the higher order
contact terms), respectively.


For each partial wave $l$ unitarity imposes a restriction on
the (inverse) $T$-matrix above the pertinent thresholds
\beq \label{unit}
\mbox{Im} T^{-1}_l = - \frac{|\mathbf{q}_{cm}|}{8 \pi \sqrt{s}}
\eeq
with $\mathbf{q}_{cm}$ being the three-momentum in the
center-of-mass frame of the channel under consideration.
Since we are primarily concerned with a narrow center-of-mass energy region around
the $\bar{K} N$ threshold, it is sufficient to restrict ourselves to the
$s$-wave (matrix) amplitude $V(s)$.
We write the inverse of the $T$-matrix as
(suppressing the subscript $l\ ( = 0)$ for brevity)
\beq \label{invers}
T^{-1} = V^{-1} + G \ ,
\eeq
which yields after inversion
\beq \label{V}
T = [1 + V \cdot G]^{-1} \; V .
\eeq
The quantity $G$ is the finite part of the scalar loop integral $\tilde{G}$
\beq
\tilde{G}(q^2) = \int \frac{d^d p}{(2 \pi)^d}
\frac{i}{[ (q-p)^2 - M_B^2 + i \epsilon]
   [ p^2 - m_\phi^2 + i \epsilon] } ~ ,
\eeq
where $M_B$ and $m_\phi$ are the physical masses of
the baryon and the meson, respectively.
In dimensional regularization one obtains for the finite part 
$G$ of $\tilde{G}$ \cite{OM, BNW1}, 
\beqa
G(q^2) & = & a({\mu}) + \frac{1}{32 \pi^2 q^2} \Bigg\{ q^2
             \left[ \ln\Big(\frac{m_\phi^2 }{\mu^2}\Big) +
             \ln\Big(\frac{M_B^2 }{\mu^2}\Big) -2 \right]\no \\
       &   & + (m_\phi^2 - M_B^2)  \ln\left(\frac{m_\phi^2 }{M_B^2}\right)
             - 8 \sqrt{q^2} \,|\mathbf{q}_{cm}| \ \mbox{artanh}
             \left(\frac{2 \sqrt{q^2} \ |\mathbf{q}_{\scriptstyle{cm}}|}{
             (m_\phi + M_B )^2 - q^2} \right) \Bigg\}\ ,
\eeqa
where $\mu$ is the regularization scale.
The subtraction constant $a({\mu})$ cancels the scale dependence of
the chiral logarithms and simulates higher order contributions with
the value of $a({\mu})$ depending on the respective channel. Stated 
differently, this introduces some additional SU(3) breaking beyond the
use of the physical masses in the kinematics and loop functions.
Eq.~(\ref{V}) is a matrix equation with the diagonal matrix $G$
collecting the loop integrals in each channel.
This amounts to a summation of a bubble chain to all orders in the $s$-channel,
equivalent to solving a Bethe-Salpeter equation with $V$ as driving term,
where all momenta in $V$ are set to their on-shell values. This so-called 
on-shell scheme reduces the full Bethe-Salpeter equation to the simple matrix equation 
(\ref{V}). (Note, however, that in the presence of the crossed Born term (Fig.~\ref{fig:feyns}d)  
this simplification must be treated with care due to the appearance of
unphysical subthreshold cuts \cite{BNW1}.)

Moreover, the Coulomb interaction has been shown to yield  sizable
contributions to the elastic $K^- p$ scattering amplitude up to kaon laboratory momenta of
100-150 MeV/$c$ \cite{JackDal}. We have thus taken into account  the
electromagnetic interactions as well, cf. \cite{BNW1} for details.

\section{Results}\label{sec:res}

In this section, we present our results for $K^- p$ scattering and the resulting
prediction for the $K^- p$ scattering length, $a_{K^- p}$.
Low-energy antikaon-nucleon scattering and reactions have been studied experimentally 
decades ago \cite{Hum, Sak, Kim, Kit, Eva, Cib}. The available data (admittedly with large errors) 
are mostly restricted to $K^-$ momenta above 100 MeV/c. Further tight constraints 
are imposed by the accurately determined threshold branching ratios into the inelastic channels 
$\pi \Sigma$ and $\pi^0 \Lambda$ \cite{Now, Tov}:
\beqa \label{eq:BRdef}
\gamma & = & \frac{\Gamma(K^- p \to \pi^+ \Sigma^-)}{\Gamma(K^- p \to \pi^- \Sigma^+)}
         = 2.36 \pm 0.04 , \no \\
R_c & = & \frac{\Gamma(K^- p \to \pi^+ \Sigma^- , \ \pi^- \Sigma^+)}
          {\Gamma(K^- p \to \textrm{\small all inelastic channels})}
      = 0.664 \pm 0.011 , \no \\
R_n & = & \frac{\Gamma(K^- p \to \pi^0 \Lambda)}{\Gamma(K^- p \to \textrm{\small neutral states})}
      = 0.189 \pm 0.015\, ,
\eeqa
and by the $\pi \Sigma$ invariant mass spectrum in the isospin $I = 0$ channel \cite{Hem}.

In our analysis we have restricted ourselves to pure meson-baryon scattering and have not included the 
processes $K^- p \to \gamma \Lambda(1405)$
\cite{NOTR}, $\gamma p \to K^* \Lambda(1405)$ \cite{HHVO} which is now experimentally
under investigation at Spring8/Osaka \cite{Spring8} and, in the near future,
also at ELSA (Bonn), and 
$K^- p \to \pi^0 \pi^0 \Sigma^0$ \cite{MOR} which is already measured \cite{CrBa}.
Reactions including the coupling to an external photon such as $K^- p \to \gamma \Lambda(1405)$ and
$\gamma p \to K^* \Lambda(1405)$ require substantial extension of the chiral unitary approach
applied here as illustrated, e.g., in \cite{BBMN}, whereas the three-body
final state in $K^- p \to \pi^0 \pi^0 \Sigma^0$ introduces additional 
model-dependence. Hence, these processes are beyond the scope of the present investigation.

Our approaches have six subtraction constants $a(\mu)$ in the different channels
and up to eight parameters which are varied in the fits within generous limits: 
the decay constant $f$ and in the full approach the higher order couplings $b_i$, $d_i$.
The axial vector couplings $D$, $F$ which enter the Born diagrams (Fig.~\ref{fig:feyns}b and c)
are kept fixed at the values $D = 0.80$, $F = 0.46$ extracted from semileptonic hyperon decays 
\cite{CR}.
We have purposely chosen ample ranges for the parameters of our approach, in order
to be able to take into account a large variety of qualitatively different fits
to $K^- p$ scattering data.
Since our concern here is to predict the strong
$K^- p$ scattering length only from $K^- p$ scattering data, we do not impose additional 
phenomenological constraints, e.g., the analysis of \cite{BMW} which includes 
$\eta$ photoproduction as a high quality data set. 
In addition, the framework chosen in \cite{BMW} does not exactly coincide with any of 
the approaches considered in the present work. We can therefore not expect the same 
values for the coupling constants.

We perform an overall least-squares fit to available low-energy $K^- p$ scattering data
for the three different approaches, ``WT'', ``WTB'' and ``full''. 
To this end, we first calculate the individual $\chi_i^2$ for the $i$-th
observable and divide by the number of pertinent data points $n_i$.
The total $\chi^2$ per degree of freedom (d.o.f.) is then defined as \cite{Hoehler}
\beq \label{eq:chi^2}
\frac{\chi^2}{\mbox{d.o.f.}} = \frac{\sum_i n_i}{N (\sum_i n_i - p)}  \sum_i \frac{\chi_i^2}{n_i} \ ,
\eeq
where $N$ is the number of observables and $p$ the number of free parameters in the approach.
This definition of the $\chi^2/\mbox{d.o.f.}$ function generalizes the standard
$\chi^2/\mbox{d.o.f.}$ for a single observable and has the advantage that all observables 
are weighted equally regardless of the number of data points. 
If one were to use instead the definition $\chi^2/\mbox{d.o.f.}= \sum_i \chi_i^2/(\sum_i n_i - p)$
in which all data points from different observables have the same weight then
single-valued observables (such as branching ratios) would be dominated by observables 
with many data points (such as scattering data). Note that the definition in Eq.~(\ref{eq:chi^2}) 
reduces to the latter one if all observables have the same number of data points.

With respect to the work \cite{BNW1} we have significantly improved
our fitting procedure which now involves the combination of a Monte Carlo
routine with a conjugate gradient method \cite{GSL}. 
This allows us to perform a large number of different fits to 
data distributed in parameter space so that the model dependence
of the results is reduced and a realistic error range for the 
$K^- p$ scattering length derived from scattering experiments can be provided. 

We obtain a large number of fits which describe the low-energy $K^- p$ scattering data very 
well with the minimum $\chi^2/\mbox{d.o.f.}$ values given by 1.28, 0.88 and 0.71 
for the different 
approaches ``WT'', ``WTB'' and ``full'', respectively. In general, if $\chi^2$ is a function of $n$ 
parameters $\boldsymbol{\theta} = (\theta_1, \dots, \theta_n)$ the standard error range of 
these parameters is given by the condition \cite{pdg}
\beq \label{eq:confreg}
\chi^2(\boldsymbol{\theta}) = \chi^2_{\textrm{min}} + \Delta\chi^2
\eeq
and $\Delta\chi^2$ is derived from the $p$-value of the $\chi^2$ probability distribution function
with the pertinent number of degrees of freedom. Strictly speaking, this relation only 
holds if the method of least squares is applied to one single experiment and the associated 
fit function depends linearly on the parameters $\boldsymbol{\theta}$. In the present investigation
the situation is more involved: the free parameters of the approach enter in a highly 
non-linear way into the calculation of observables and the $\chi^2$-function which is minimized
combines a variety of measurements of different quantities. Nevertheless, we will adopt the 
standard definition of a confidence region, Eq.~(\ref{eq:confreg}), since 
one can expect it to be a reasonable approximation---at least in the vicinity of the minimum of the
$\chi^2$-function, where its shape should be nearly parabolic. Our fit includes a total number of 
171 data points and either 7 (``WT'', ``WTB'') or 14 (``full'') parameters. The
region of one standard deviation (i.e.\ 68.27\% confidence level) is then found by adding
$\Delta\chi^2/\mbox{d.o.f.} = 1.05$ to the minimal $\chi^2/\mbox{d.o.f.}$ as defined 
in Eq.~(\ref{eq:chi^2}) regardless of the approach (``WT'', ``WTB'' or ``full'') since the difference 
in the number of parameters causes only tiny modifications in $\Delta\chi^2/\mbox{d.o.f.}$.

As it turns out, we also obtain fits that although having a relatively low overall 
$\chi^2/\mbox{d.o.f.}$ fail
miserably in one or two observables and can thus not be classified as ``good fits''. In addition 
to the bound on the overall $\chi^2/\mbox{d.o.f.}$ we thus demand that each individual observable be
reproduced with a $\chi_i^2/n_i$ value which does not exceed the overall $\chi^2/\mbox{d.o.f.}$ by more 
than a factor of four. 
Fits which do not meet this additional criterion are grouped instead with fits with
the lowest $\chi^2/\mbox{d.o.f.}$ value which satisfies this constraint.
This specific choice has proved useful in practice.
These two goodness-of-fit criteria, a bound on the overall $\chi^2$ and on the individual
$\chi_i^2$, determine the error regions specified in the following for all parameters and observables,
while the presented central values correspond to the fits with minimal $\chi^2$.
We also investigate analytic properties of the fits, in order
to sort out solutions with unphysical pole structures. This issue will be discussed 
in detail in the next section.

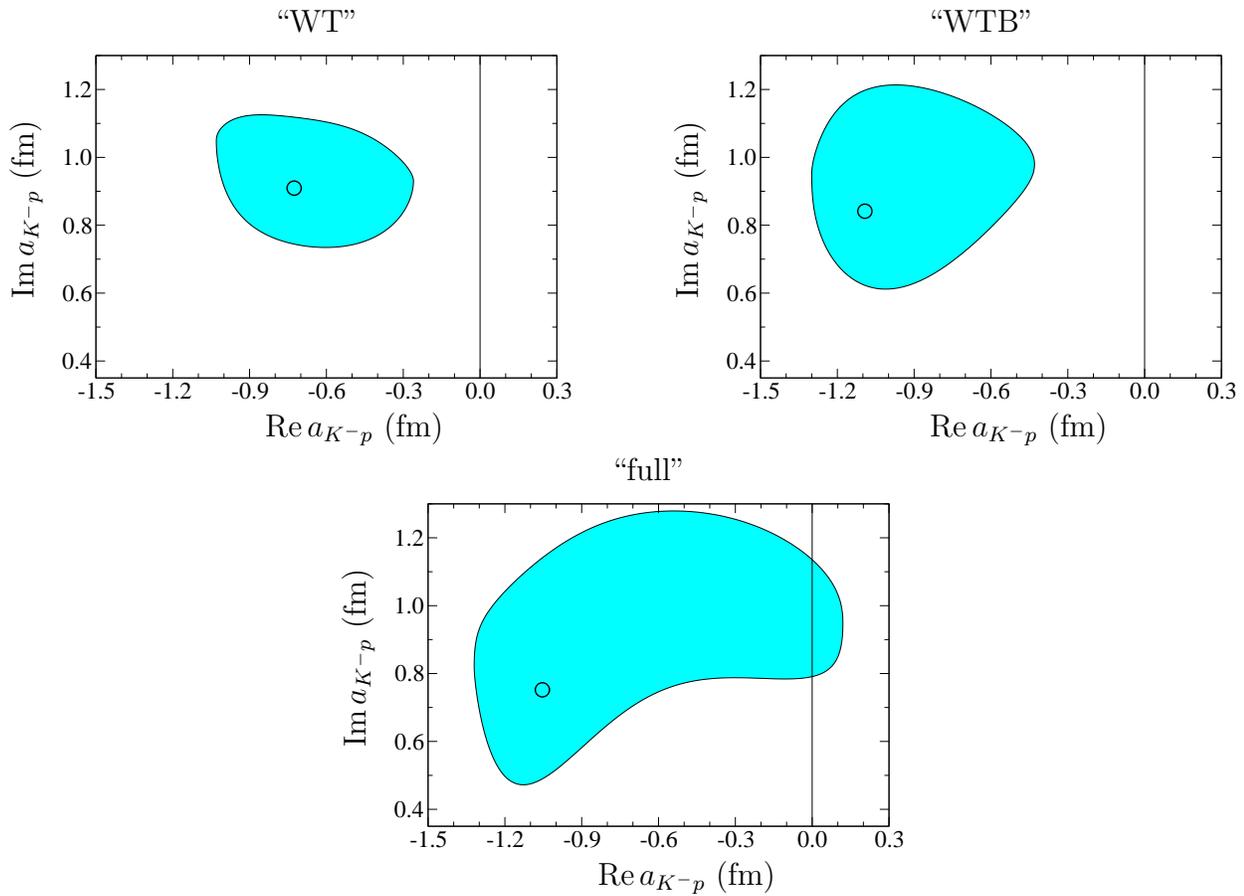
\begin{figure}
\centering
\begin{tabular}{ccc}
``WT'' & & ``WTB'' \\[1ex]
\begin{overpic}[width=0.4\textwidth,clip]{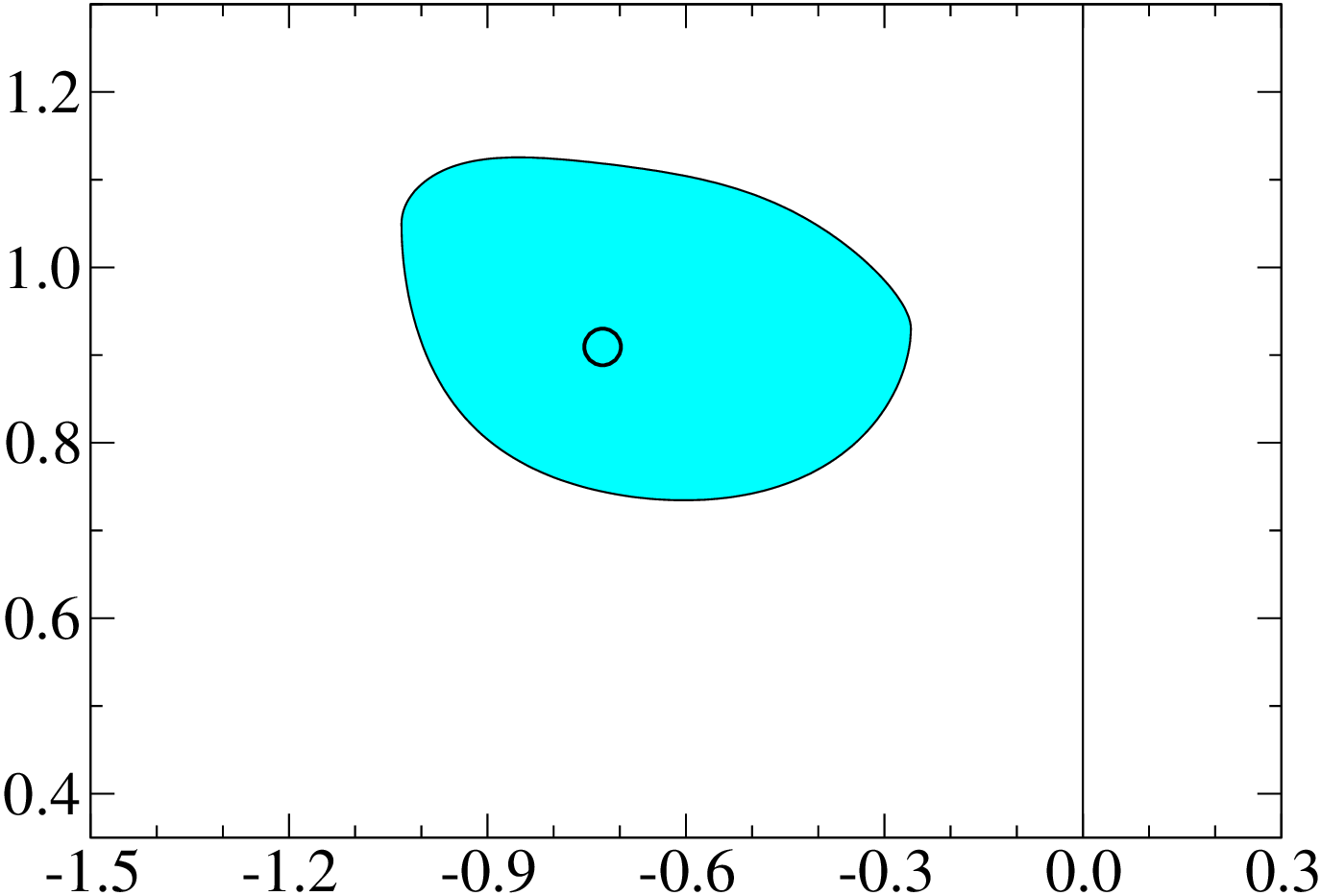}
\put(40,-7){\scalebox{1.0}{$\Re{a_{K^- p}}$ (fm)}}
\put(-10,20){\rotatebox{90}{\scalebox{1.0}{$\Im{a_{K^- p}}$ (fm)}}}
\end{overpic}
& \hspace{0.07\textwidth} &
\begin{overpic}[width=0.4\textwidth,clip]{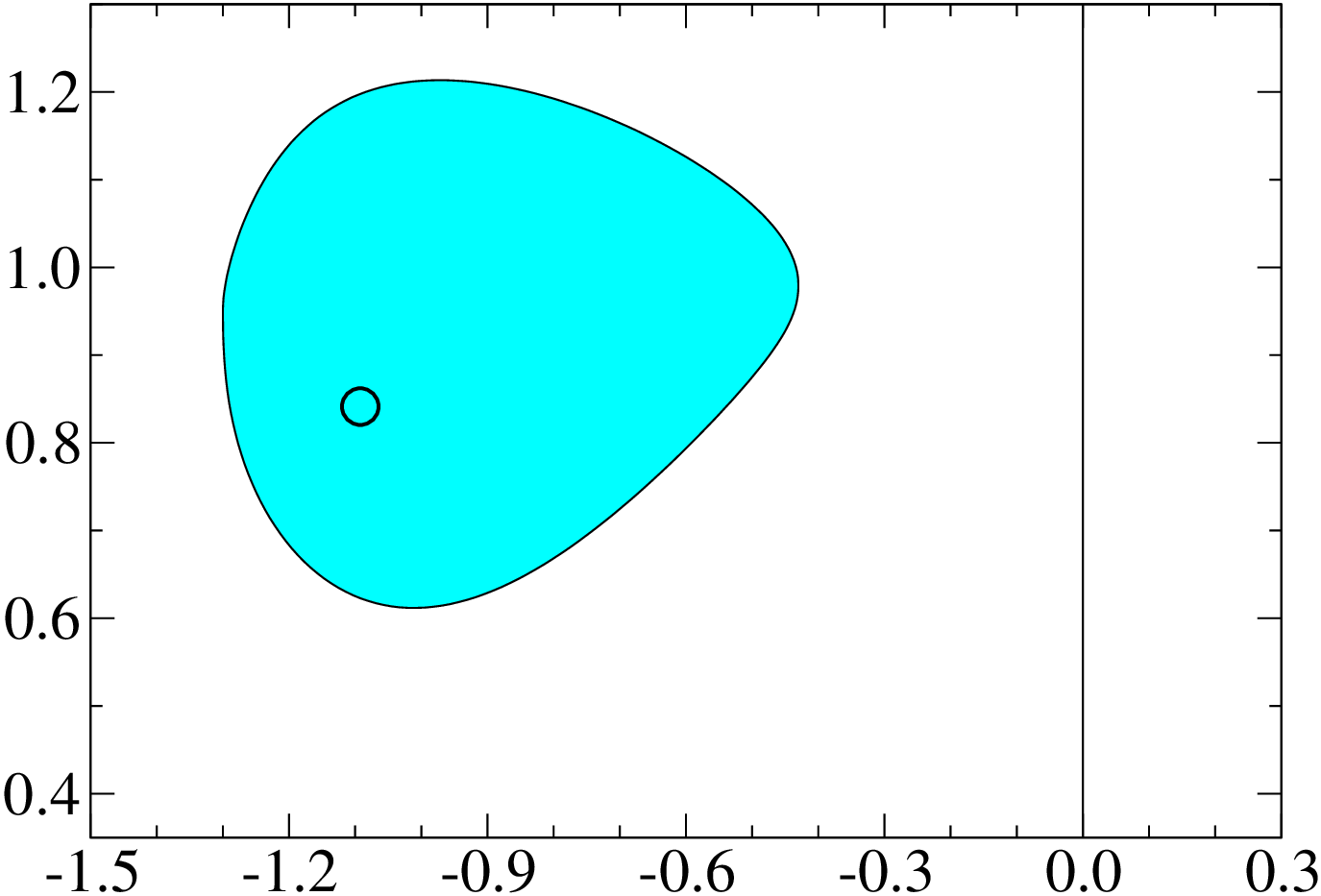}
\put(40,-7){\scalebox{1.0}{$\Re{a_{K^- p}}$ (fm)}}
\put(-10,20){\rotatebox{90}{\scalebox{1.0}{$\Im{a_{K^- p}}$ (fm)}}}
\end{overpic} \\[3ex]
\multicolumn{3}{c}{``full''}\\[1ex]
\multicolumn{3}{c}{%
\begin{overpic}[width=0.4\textwidth,clip]{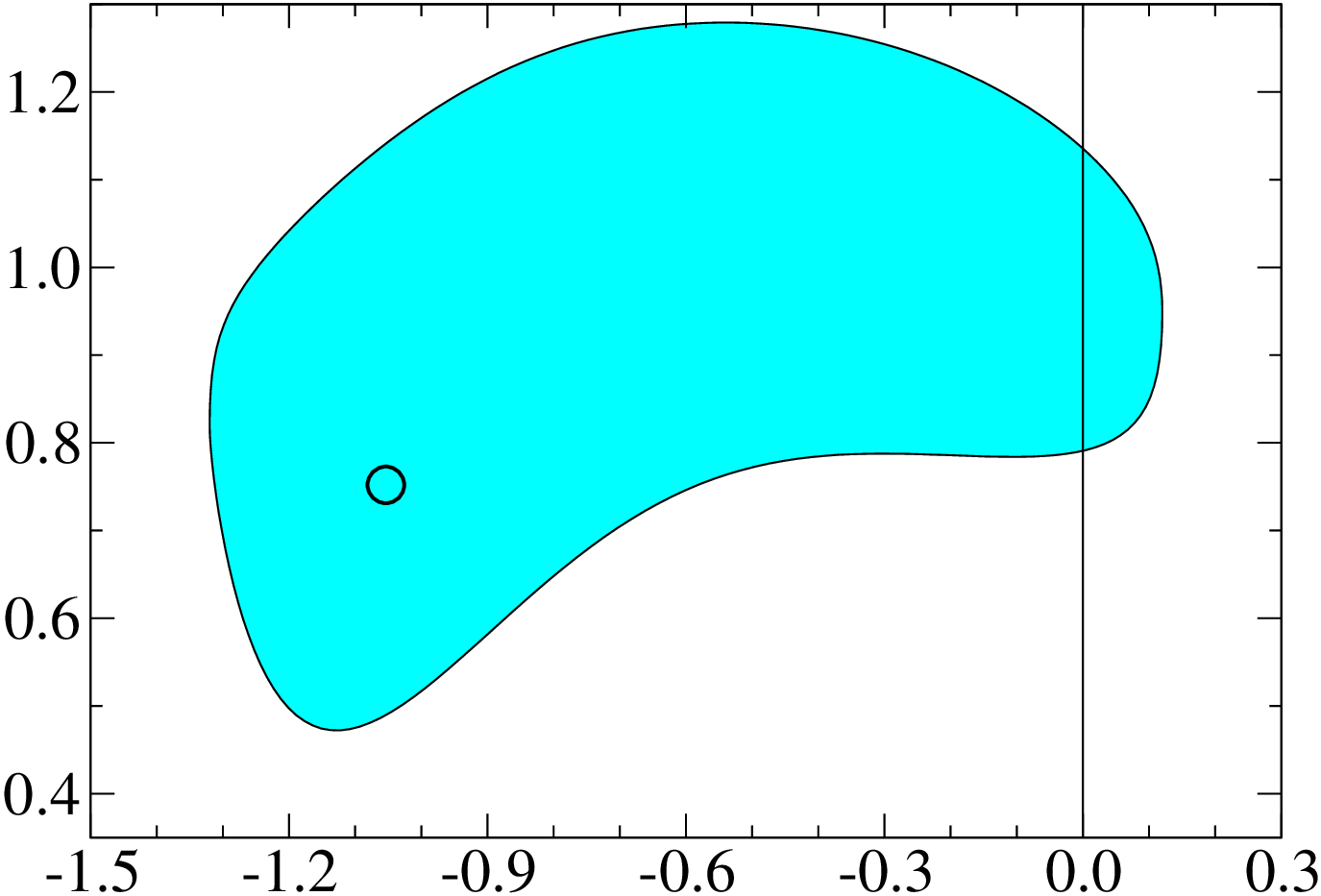}
\put(40,-7){\scalebox{1.0}{$\Re{a_{K^- p}}$ (fm)}}
\put(-10,20){\rotatebox{90}{\scalebox{1.0}{$\Im{a_{K^- p}}$ (fm)}}}
\end{overpic}}\\[2ex]
\end{tabular}
\caption{Real and imaginary parts of the scattering length $a_{K^- p}$ for the three approaches.
         The circles indicate the result of the best fits, the shaded areas represent the 
         1$\sigma$ confidence region.}
\label{fig:aKmp}
\end{figure}

The scattering length $a_{K^- p}$ is given by the strong interaction $T$ matrix at threshold
\beq
a_{K^- p} = \frac{1}{8 \pi \sqrt{s}} \ T_{K^- p \to K^- p}(s) \Big|_{s = (m_{K^-} + M_p)^2} \quad .
\eeq
The values corresponding to the best fits in the three approaches are
\beq
\begin{array} {ll}
\textrm{``WT'':}   \quad & a_{K^- p} = (-0.73 + i\,0.91) \fm  \ , \\
\textrm{``WTB'':}  \quad & a_{K^- p} = (-1.09 + i\,0.84) \fm  \ , \\
\textrm{``full'':} \quad & a_{K^- p} = (-1.05 + i\,0.75) \fm  \ . \\
\end{array}
\eeq
The errors of the real and imaginary parts are, of course, correlated and the
1$\sigma$ regions in the complex $a_{K^- p}$ plane are depicted in Fig.~\ref{fig:aKmp}.
While the absolute minimum of $\chi^2/\mbox{d.o.f.}$ is the lowest in the full
approach, the $\chi^2$ function rises steeper in the ``WT'' and ``WTB'' approaches leading 
to a smaller 1$\sigma$ confidence region in the $a_{K^- p}$ plane.

\begin{figure}
\centering
\begin{tabular}{ccc}
``WT'' & & ``WTB'' \\[1ex]
\begin{overpic}[width=0.4\textwidth,clip]{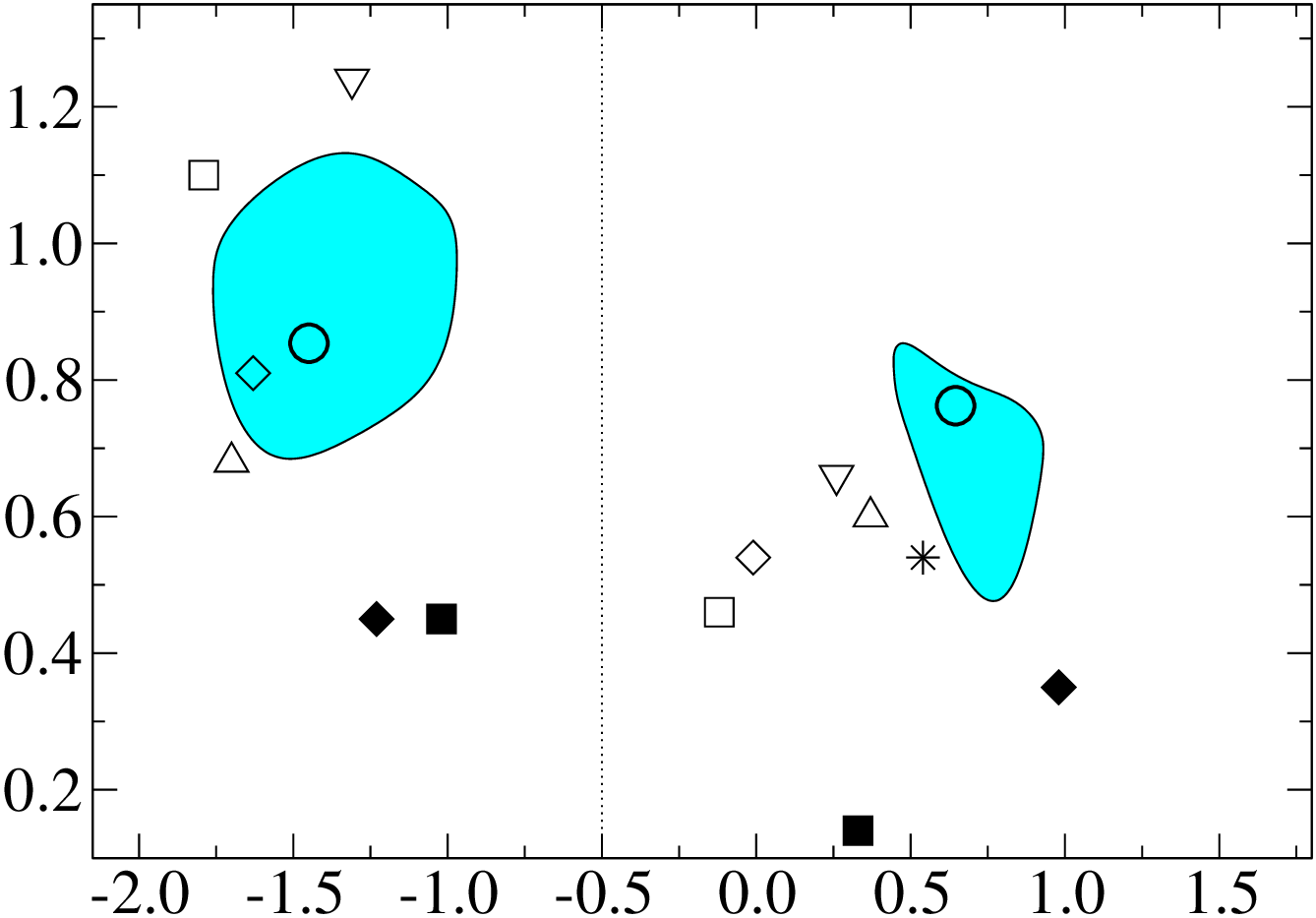}
\put(40,-7){\scalebox{1.0}{$\Re{a_i}$ (fm)}}
\put(-10,25){\rotatebox{90}{\scalebox{1.0}{$\Im{a_i}$ (fm)}}}
\put(27,42){\scalebox{1.2}{$a_0$}}
\put(72,30){\scalebox{1.2}{$a_1$}}
\end{overpic}
& \hspace{0.07\textwidth} &
\begin{overpic}[width=0.4\textwidth,clip]{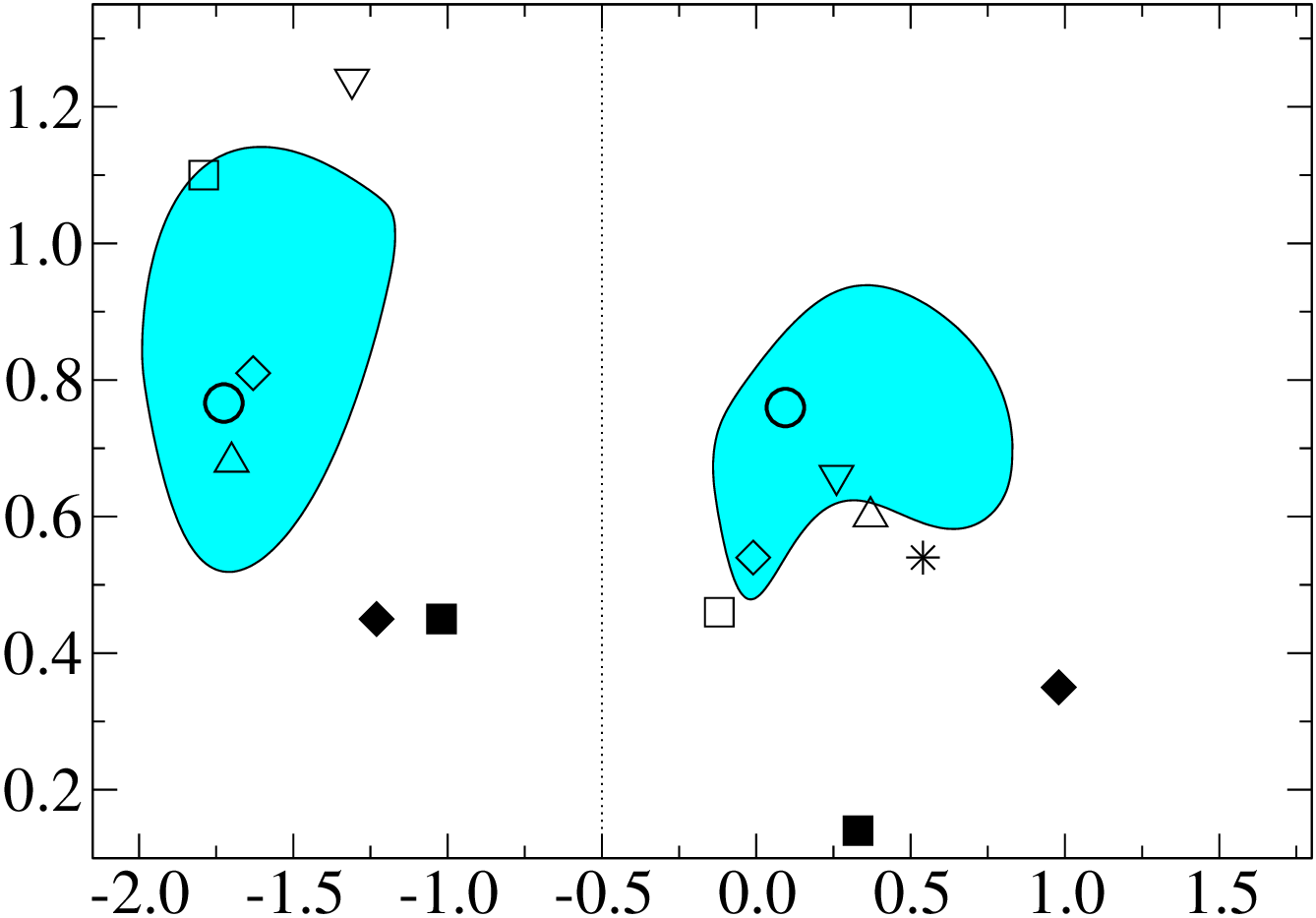}
\put(40,-7){\scalebox{1.0}{$\Re{a_i}$ (fm)}}
\put(-10,25){\rotatebox{90}{\scalebox{1.0}{$\Im{a_i}$ (fm)}}}
\put(21,46){\scalebox{1.2}{$a_0$}}
\put(64,40){\scalebox{1.2}{$a_1$}}
\end{overpic} \\[3ex]
\multicolumn{3}{c}{``full''}\\[1ex]
\multicolumn{3}{c}{%
\begin{overpic}[width=0.4\textwidth,clip]{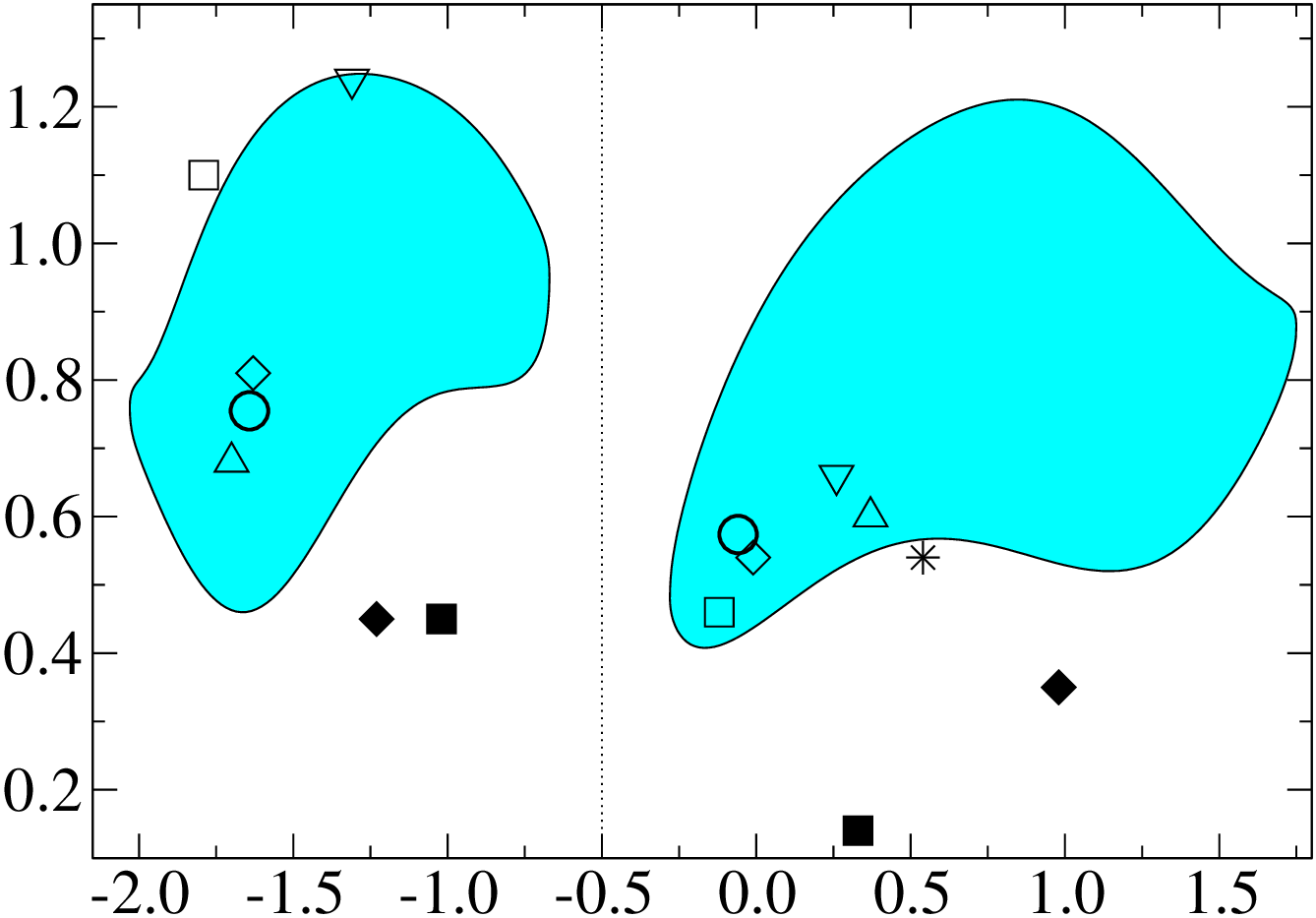}
\put(40,-7){\scalebox{1.0}{$\Re{a_i}$ (fm)}}
\put(-10,25){\rotatebox{90}{\scalebox{1.0}{$\Im{a_i}$ (fm)}}}
\put(27,44){\scalebox{1.2}{$a_0$}}
\put(74,44){\scalebox{1.2}{$a_1$}}
\end{overpic}}\\[2ex]
\end{tabular}
\caption{Real and imaginary parts of the isospin $\bar{K} N$ scattering lengths $a_0$ and $a_1$ 
         for the three approaches. The circles indicate the result of the best fits, the shaded 
         areas represent the 1$\sigma$ confidence regions.
         Our results are compared to the values found in \cite{ADM} (triangle up), \cite{OR} (star, 
         the value $a_0 = (-2.24 + 1.94i)\fm$ is not shown), \cite{OM} (triangle down), \cite{OPV}
         (filled diamond for fit $A_4^+$, empty diamond for fit $B_4^+$), \cite{Oller}
         (filled square for fit I, empty square for fit II).}
\label{fig:a0a1}
\end{figure}
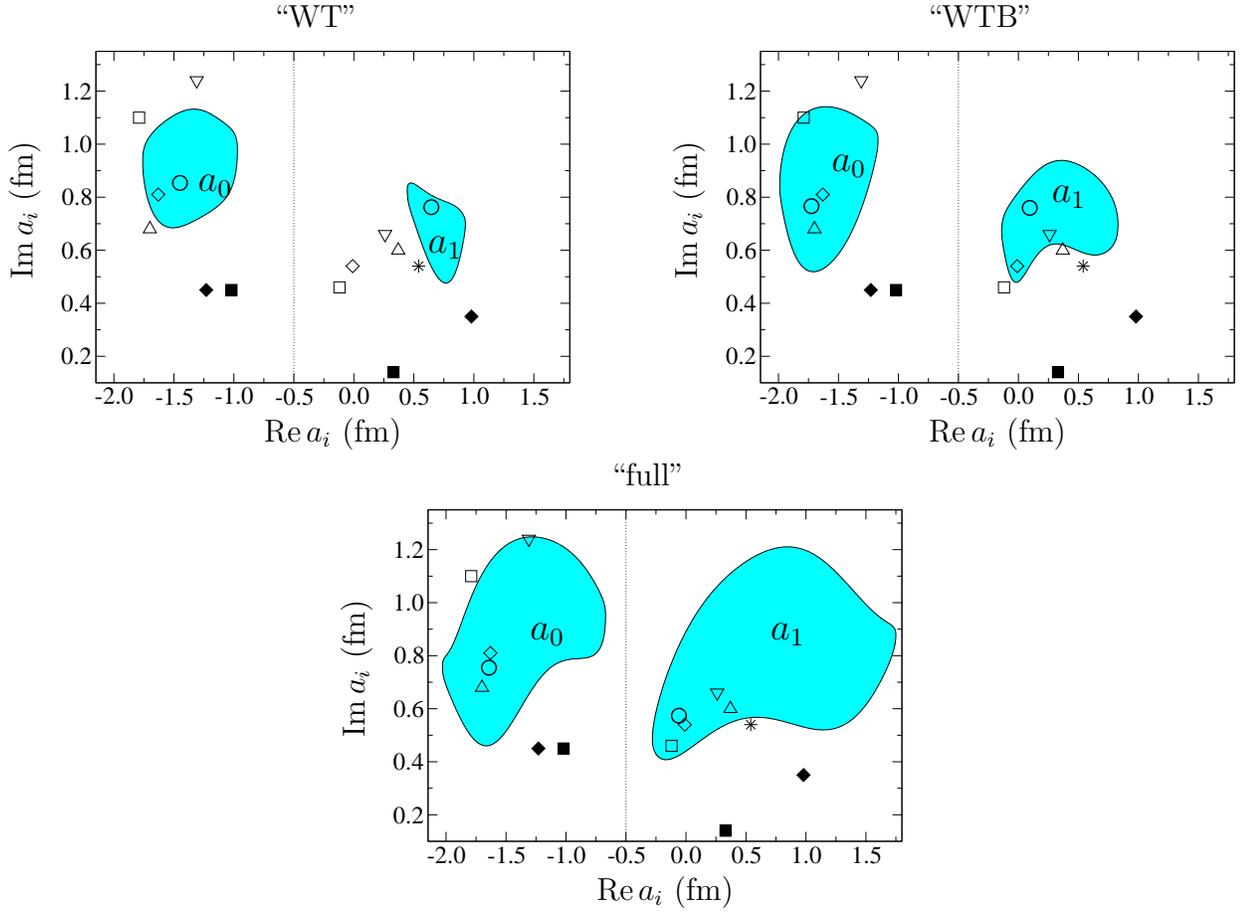

We also extract the $\bar{K}N$ $s$-wave scattering lengths $a_0$, $a_1$ in the isospin limit of
equal up- and down-quark masses. Since we neglect isospin-breaking corrections in the Lagrangian
from which the interaction kernel of the coupled-channels calculation is derived, taking the isospin
limit
amounts to replacing the physical masses of the particles that enter all kinematic quantities
and the scalar loop integrals $G$ by a common mass for each isospin multiplet
and we disregard any electromagnetic effect.
For the different approaches we obtain the central values
\beq
\begin{array} {lll}
\textrm{``WT'':}   \quad & a_0 = (-1.45 + i\,0.85) \fm  \ , \quad & a_1 = ( 0.65 + i\,0.76)\fm \ , \\
\textrm{``WTB'':}  \quad & a_0 = (-1.72 + i\,0.77) \fm  \ , \quad & a_1 = ( 0.09 + i\,0.76)\fm \ , \\
\textrm{``full'':} \quad & a_0 = (-1.64 + i\,0.75) \fm  \ , \quad & a_1 = (-0.06 + i\,0.57)\fm \ . \\
\end{array}
\eeq
In Fig.~\ref{fig:a0a1} we show the error ranges and compare the results with values found 
in similar chiral unitary approaches \cite{OR, OM, OPV, Oller} and a multichannel dispersion relation
analysis of $\bar{K} N$ scattering data \cite{ADM}. 
One observes consistency of all three approaches (``WT'', ``WTB'', ``full'') within error bars
and agreement with most of the values from previous investigations. 
However, our results do not agree with fit $A_4^+$ in \cite{OPV} (as already discussed in \cite{BNW2})
and the similar fit I in \cite{Oller}. They also disagree with the $a_1$ value
in \cite{OR}, where a variant of the ``WT'' approach was employed.
The approach utilized in \cite{OM} nearly coincides with ``WTB'' in the present work, 
however only one common subtraction constant for all channels was employed in \cite{OM}, while 
in the present work we have the freedom to vary six (isospin symmetric) subtraction constants.
This explains why the imaginary part of $a_0$ in \cite{OM} is larger and outside 
the 1$\sigma$ range of the present calculation.

Due to the very high statistics of the present investigation our results provide a realistic 
error range for the isospin scattering lengths $a_0$ and $a_1$ within chiral unitary approaches. 
As shown in \cite{MRR2} these quantities are an 
important input in the theoretical analysis of the upcoming spectroscopy study of
kaonic deuterium \cite{SID} and further anticipated experiments with even more complex light 
kaonic nuclei \cite{AMA} at DA$\Phi$NE.

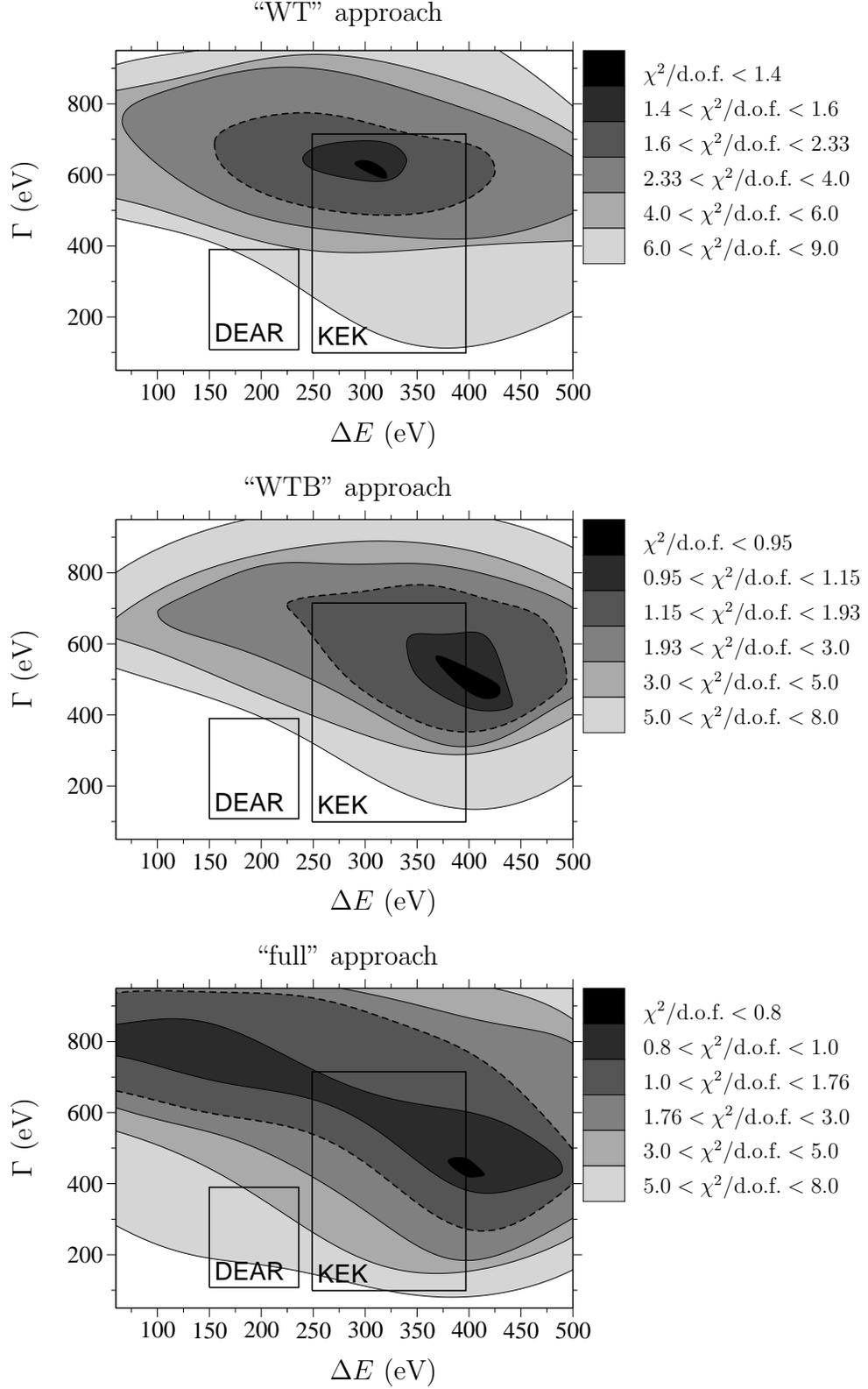
\begin{figure}
\centering
\begin{tabular}{c}
``WT'' approach \\[1ex]
\begin{overpic}[width=0.5\textwidth,clip]{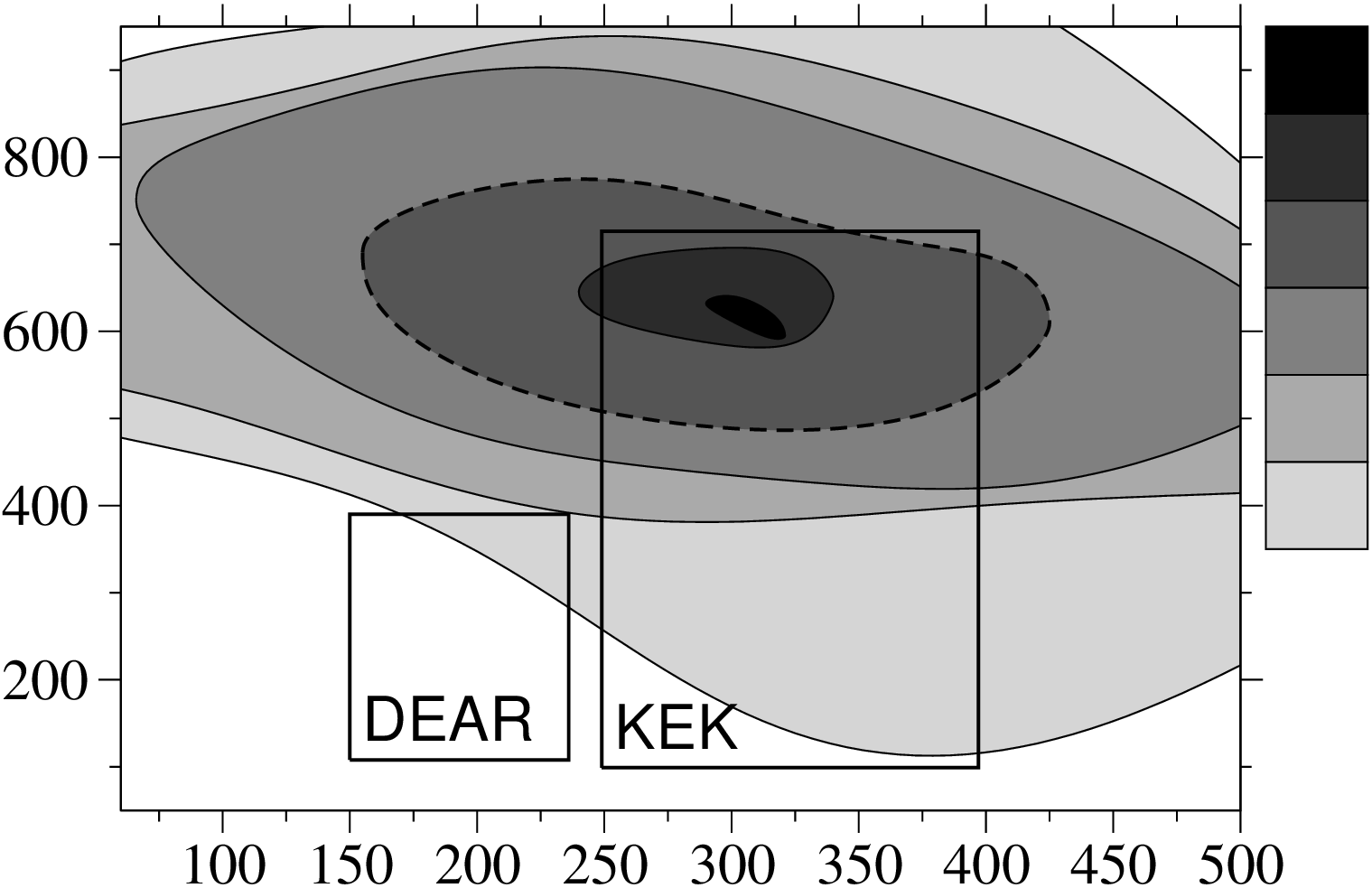}
\put(47,-7){\scalebox{1.0}{$\Delta E$ (eV)}}
\put(-10,29){\rotatebox{90}{\scalebox{1.0}{$\Gamma$ (eV)}}}
\put(103,57.3){\scalebox{0.8}{$\chi^2/\textrm{d.o.f.} < 1.4$}}
\put(103,51.08){\scalebox{0.8}{$1.4 < \chi^2/\textrm{d.o.f.} < 1.6$}}
\put(103,44.86){\scalebox{0.8}{$1.6 < \chi^2/\textrm{d.o.f.} < 2.33$}}
\put(103,38.64){\scalebox{0.8}{$2.33 < \chi^2/\textrm{d.o.f.} < 4.0$}}
\put(103,32.42){\scalebox{0.8}{$4.0 < \chi^2/\textrm{d.o.f.} < 6.0$}}
\put(103,26.2){\scalebox{0.8}{$6.0 < \chi^2/\textrm{d.o.f.} < 9.0$}}
\end{overpic}\\[5ex]
``WTB'' approach \\[0.5ex]
\begin{overpic}[width=0.5\textwidth,clip]{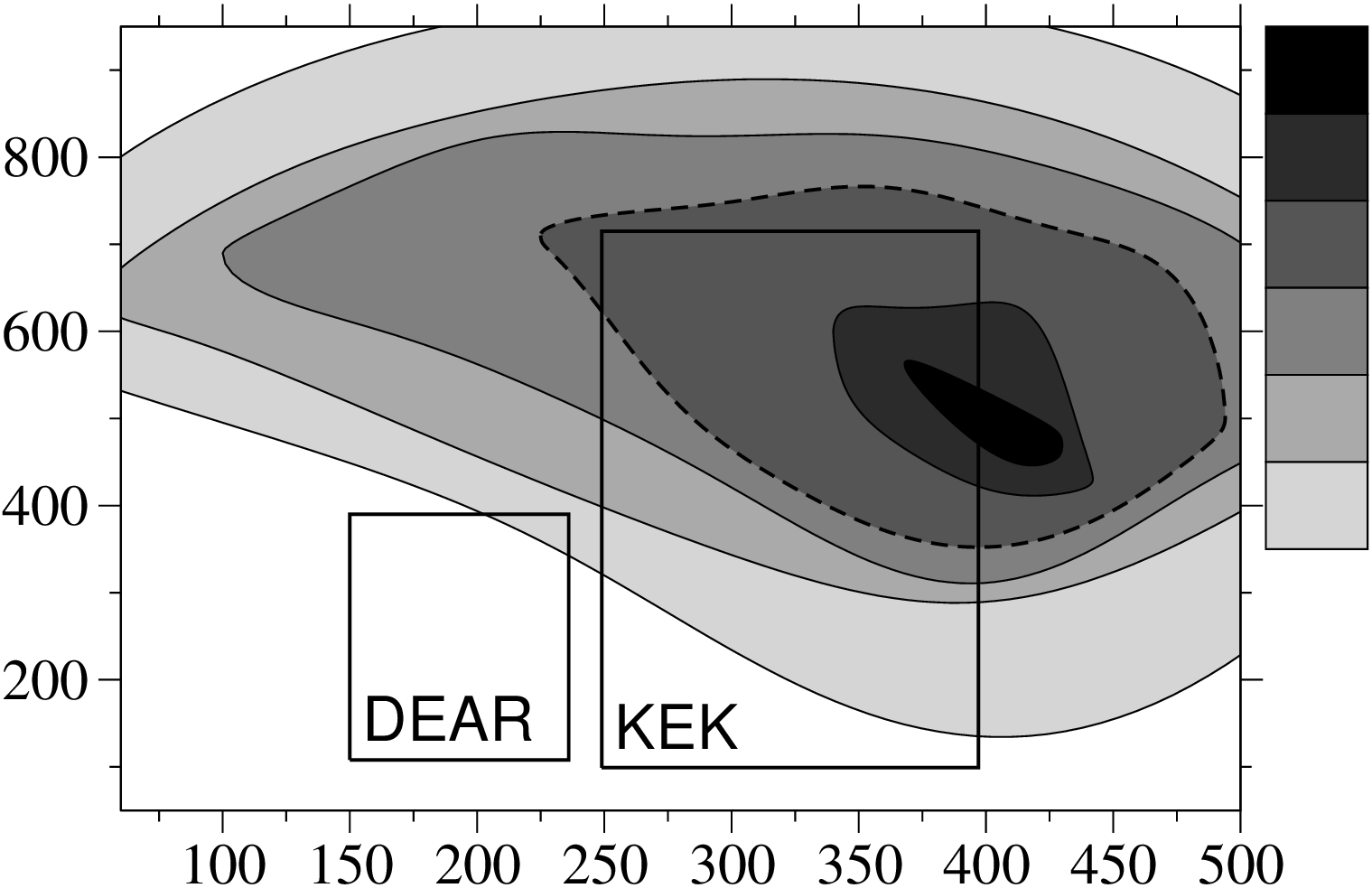}
\put(47,-7){\scalebox{1.0}{$\Delta E$ (eV)}}
\put(-10,29){\rotatebox{90}{\scalebox{1.0}{$\Gamma$ (eV)}}}
\put(103,57.3){\scalebox{0.8}{$\chi^2/\textrm{d.o.f.} < 0.95$}}
\put(103,51.08){\scalebox{0.8}{$0.95 < \chi^2/\textrm{d.o.f.} < 1.15$}}
\put(103,44.86){\scalebox{0.8}{$1.15 < \chi^2/\textrm{d.o.f.} < 1.93$}}
\put(103,38.64){\scalebox{0.8}{$1.93 < \chi^2/\textrm{d.o.f.} < 3.0$}}
\put(103,32.42){\scalebox{0.8}{$3.0 < \chi^2/\textrm{d.o.f.} < 5.0$}}
\put(103,26.2){\scalebox{0.8}{$5.0 < \chi^2/\textrm{d.o.f.} < 8.0$}}
\end{overpic} \\[5ex]
``full'' approach \\[0.5ex]
\begin{overpic}[width=0.5\textwidth,clip]{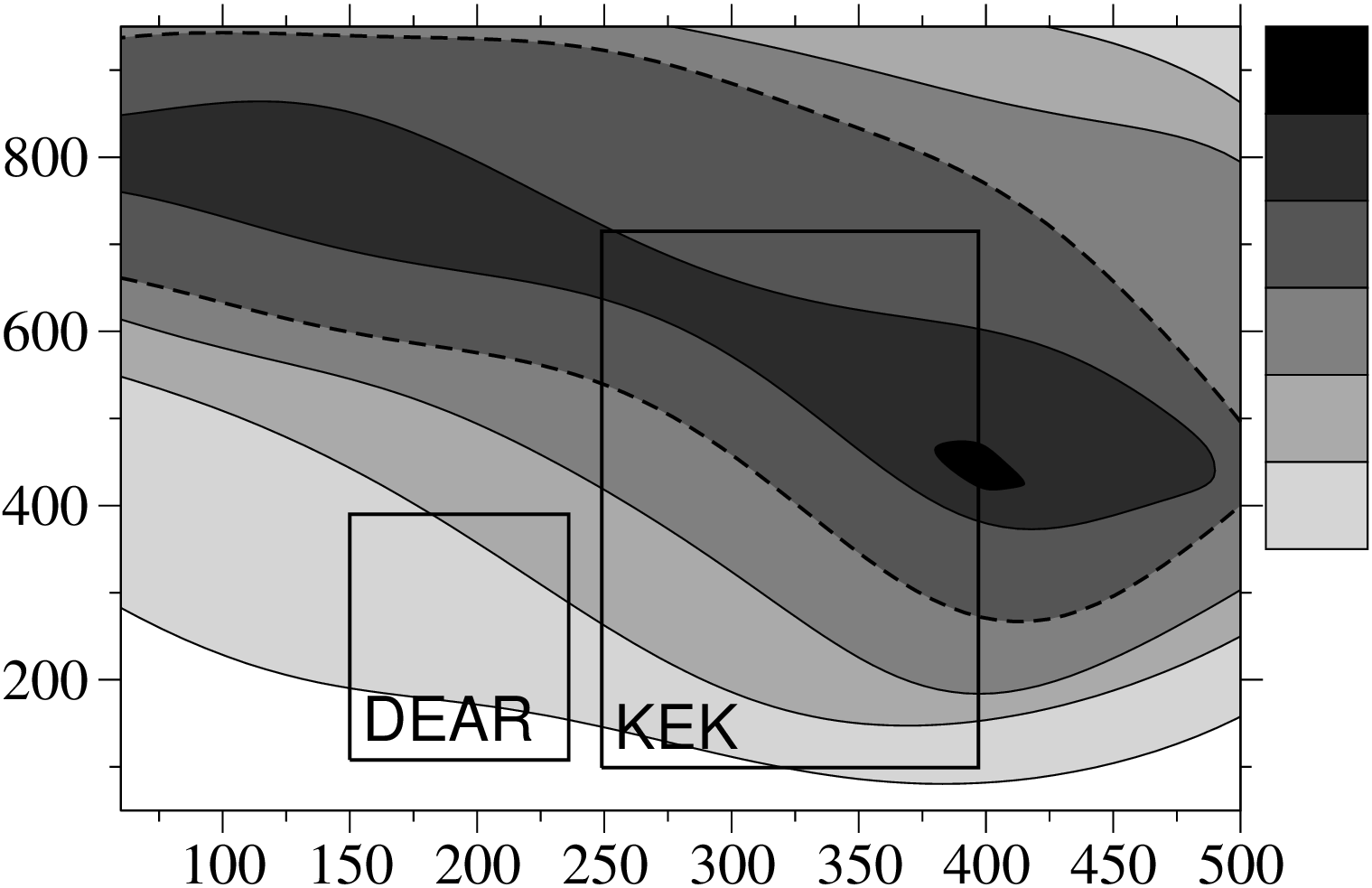}
\put(47,-7){\scalebox{1.0}{$\Delta E$ (eV)}}
\put(-10,29){\rotatebox{90}{\scalebox{1.0}{$\Gamma$ (eV)}}}
\put(103,57.3){\scalebox{0.8}{$\chi^2/\textrm{d.o.f.} < 0.8$}}
\put(103,51.08){\scalebox{0.8}{$0.8 < \chi^2/\textrm{d.o.f.} < 1.0$}}
\put(103,44.86){\scalebox{0.8}{$1.0 < \chi^2/\textrm{d.o.f.} < 1.76$}}
\put(103,38.64){\scalebox{0.8}{$1.76 < \chi^2/\textrm{d.o.f.} < 3.0$}}
\put(103,32.42){\scalebox{0.8}{$3.0 < \chi^2/\textrm{d.o.f.} < 5.0$}}
\put(103,26.2){\scalebox{0.8}{$5.0 < \chi^2/\textrm{d.o.f.} < 8.0$}}
\end{overpic}\\[2ex]
\end{tabular}
\caption{Strong energy shift $\Delta E$ and width $\Gamma$ of kaonic hydrogen for the three approaches.
         The shaded areas represent different upper limits of the overall $\chi^2/\mbox{d.o.f.}$\@
         The 1$\sigma$ confidence region is bordered by the dashed line. 
         See text for further details.}
\label{fig:KHyd}
\end{figure}

In the presence of electromagnetic corrections
the ground state strong energy shift $\Delta E$ and width $\Gamma$
of kaonic hydrogen are related to the $K^- p$ scattering length $a_{K^- p}$ via \cite{MRR}
\beq \label{eq:Rus}
\Delta E - \frac{i}{2} \Gamma = -2 \alpha^3 \mu_{c}^2 a_{K^- p} \ 
    [1 - 2 \alpha \mu_{c} (\ln{\alpha} -1) a_{K^- p}] \ ,
\eeq
where $\mu_c$ is the reduced mass of the $K^- p$ system and $\alpha$ is the 
fine-structure constant.
The obtained predictions for $\Delta E$ and $\Gamma$,
which are solely based on $K^- p$ scattering data and the $\pi \Sigma$ invariant mass spectrum,
are presented in Fig.~\ref{fig:KHyd} for the three different approaches. They are compared to the recent 
experimental determination of kaonic hydrogen observables by both 
the KEK \cite{KEK} and the DEAR \cite{DEAR}
collaborations. Figure~\ref{fig:KHyd} constitutes together with Figs.~\ref{fig:aKmp} and \ref{fig:a0a1} 
the main result of the present work.
The shaded areas in the plots of Fig.\ref{fig:KHyd}
represent smoothened areas which correspond to different upper limits of the 
overall $\chi^2/\mbox{d.o.f.}$ (with the additional constraint that each individual observable is 
reproduced by the fit with $\chi^2_i/n_i$ of at most four times the upper limit of $\chi^2/\mbox{d.o.f.}$)
and are drawn on the basis of several thousands of fits (e.g.\ more than 7000 for the full approach).
We point out that a comparable statistical exploration of parameter space in chiral unitary
approaches for $\bar{K} N$ interactions has not been attempted before; it provides for the first time
a realistic estimate of theoretical uncertainties within this framework.

Regardless of the chosen approach the fits with minimal overall
$\chi^2$ agree nicely with the result of the KEK experiment, while the 1$\sigma$ confidence region, which 
is bordered by the dashed 
line in the plots of Fig.~\ref{fig:KHyd}, has no overlap with the error ranges
given by the DEAR experiment. As discussed above the standard definition of the 1$\sigma$ confidence region
by means of Eq.~(\ref{eq:confreg}) is not strictly applicable in the present investigation, where the fit 
function is non-linear in the parameters and the fit incorporates a variety of different observables.
Therefore we refrain from showing 2$\sigma$ and 3$\sigma$ confidence regions since
application of the standard error estimation seems more questionable in these cases.

Instead we plot regions that correspond to quadratically increasing upper 
limits of the overall $\chi^2$. Note that fits which are compatible 
with the error ranges given by DEAR have an overall $\chi^2/\mbox{d.o.f.}$ of at least 
6.1, 5.5, 3.3 in the ``WT'', ``WTB'', ``full'' approach, respectively, 
with elastic $K^- p$ scattering being the largest source of disagreement.
Finally, for the best fit of each approach we present in Table~\ref{tab:LECs} 
the numerical values of the fitted 
low energy constants $f$, $b_0$, $b_D$, $b_F$, $d_1$, $d_2$, $d_3$, $d_4$ and the subtraction 
constants in the loop integrals $G$.
Note that in the fits we have allowed for broad ranges for the subtraction 
constants $a_{\phi B}$. In fact, in the ``WT'' and ``WTB'' approaches the resulting $a_{K \Xi}$ are roughly one order of 
magnitude larger than the remaining subtraction constants. However, the fits are not very sensitive to 
variations in this parameter such that the $\chi^2$ value is only slightly increased if
$a_{K \Xi}$ is reduced to the same size as the other $a_{\phi B}$.
The detailed comparison of experimental input from $\bar{K} N$ and $\pi \Sigma$ scattering 
with our fits is compiled in the appendix.

\begin{table}
\centering
\begin{tabular}{|l@{\,}l|r|r|r|}
\hline
& & \multicolumn{1}{c|}{``WT''} & \multicolumn{1}{c|}{``WTB''} & \multicolumn{1}{c|}{``full''} \\
\hline
\hline
$f$               &(MeV)       & $120.9$                & $ 86.0$                & $ 77.3$ \\
\hline
$b_0$             &(GeV$^{-1}$)&\multicolumn{1}{c|}{---}&\multicolumn{1}{c|}{---}& $-0.01$ \\
\hline
$b_D$             &(GeV$^{-1}$)&\multicolumn{1}{c|}{---}&\multicolumn{1}{c|}{---}& $-0.27$ \\
\hline
$b_F$             &(GeV$^{-1}$)&\multicolumn{1}{c|}{---}&\multicolumn{1}{c|}{---}& $-0.11$ \\
\hline
$d_1$             &(GeV$^{-1}$)&\multicolumn{1}{c|}{---}&\multicolumn{1}{c|}{---}& $-0.14$ \\
\hline
$d_2$             &(GeV$^{-1}$)&\multicolumn{1}{c|}{---}&\multicolumn{1}{c|}{---}& $-0.05$ \\
\hline
$d_3$             &(GeV$^{-1}$)&\multicolumn{1}{c|}{---}&\multicolumn{1}{c|}{---}& $-0.25$ \\
\hline
$d_4$             &(GeV$^{-1}$)&\multicolumn{1}{c|}{---}&\multicolumn{1}{c|}{---}& $-0.45$ \\
\hline
$a_{\bar{K} N}$   & $(10^{-3})$& $ -1.8$                & $ 1.9$                 & $ 1.0$  \\
\hline
$a_{\pi \Lambda}$ & $(10^{-3})$& $-12.4$                & $ 2.0$                 & $-6.2$  \\
\hline
$a_{\pi \Sigma}$  & $(10^{-3})$& $ -2.9$                & $ 2.4$                 & $ 1.9$  \\
\hline
$a_{\eta \Lambda}$& $(10^{-3})$& $ -1.7$                & $-0.9$                 & $-2.3$  \\
\hline
$a_{\eta \Sigma}$ & $(10^{-3})$& $ -1.4$                & $-3.7$                 & $-1.5$  \\
\hline
$a_{K \Xi}$       & $(10^{-3})$& $ 72.9$                & $20.0$                 & $-5.2$  \\
\hline
\hline
\multicolumn{2}{|l|}{$\chi^2/\textrm{d.o.f.}$}& 1.28    & 0.88                   & 0.71    \\
\hline
\end{tabular}
\caption{Numerical values of the fitted couplings and subtraction constants corresponding 
         to the best fits in the three approaches. The empirical value of the
         average meson decay constant is $f\simeq 100\,$MeV.}
\label{tab:LECs}
\end{table}

\section{Resonance poles}\label{sec:poles}

In order to cover a substantial region in parameter space we put only very loose constraints 
on the numerical values of the low energy parameters of the chiral effective Lagrangian, 
i.e., we solely fix the order of magnitude as motivated by the naturalness assumption
of couplings in the effective field theory. The subtraction constants in the 
loop integrals $G$ are permitted to vary in even larger ranges.
Starting from randomized initial values of the parameters the Monte Carlo routine utilized in 
the present investigation generates a vast number of fits and for all of them
the poles of the $T$-matrix in the complex $W = \sqrt{s}$ plane are determined. 
Although we work in the physical
basis, where isospin is broken by the physical masses of the particles in the loop integrals, 
we can classify 
the poles as being mainly of isospin $I = 0, 1, 2$ by their impact on channels 
(or channel combinations) which contain only one
isospin component (e.g.\ $\eta \Lambda$, $\eta \Sigma^0$).
The pole positions of the fits will serve as an additional constraint to rule out certain
fits which must be considered as unphysical as we will explain in the following.

While resonances are generally associated with poles in unphysical Riemann sheets of the complex
$W$ plane, the solutions should be free of poles in the physical Riemann sheet as 
required by the postulate of maximal analyticity. 
Since it is not possible to numerically explore the entire upper half-plane, we must
choose a finite region to search for poles in the physical sheet.
To this end, we dismiss all fits which exhibit a pole at a
distance of less than 250\,MeV from the real axis in the relevant energy region (1.25 -- 1.50\,GeV).
This selection criterion ensures that, even if such pathological poles were to exist, 
their influence on the real axis would be negligible. In particular, it guarantees that the
Wigner bound, which is based on causality and sets a lower limit for the derivative of the phase 
shift with respect to energy, is not violated \cite{Wig, BNW2}. Note that in a similar manner the Wigner 
condition has been employed in \cite{Pel} to constrain the numerical values of parameters in the 
context of unitarized chiral effective field theory.

Secondly, we reject fits which have a resonance pole on the relevant unphysical Riemann sheet that is 
located less than 2.5\,MeV below the real axis and 
thus corresponds to a resonance with a width of less than
5\,MeV. Such resonances would be one or two orders of 
magnitude narrower than what one would expect from the characteristic time-scale of the strong 
interactions of about $10^{-23}$\,s. Lifetimes of this order correspond to widths of several tens to 
hundreds of MeV, in agreement with typically observed hadronic resonances
in this energy region. At present, there is no 
experimental indication for an exotically long-lived stated in the energy interval under consideration
and we can safely ignore such fits.

The third item concerns the $\pi \Sigma$ event distribution \cite{Hem} which clearly shows
a peak corresponding to the $\Lambda(1405)$ resonance. We adopt the approach 
advocated in \cite{OM} which describes the experimental $\pi^- \Sigma^+$ event distribution
as originating from a generic $I=0$ source made up of unknown shares of $\pi \Sigma$ and 
$\bar{K} N$ states. As it happens we observe fits which do not exhibit a true resonance structure
around 1.4\,GeV, but merely show a broad bump that is generated by the intricate 
superposition of the two source states, $\pi \Sigma$ and $\bar{K} N$. 
As the $\pi \Sigma$ event distribution is not normalized, the normalization 
constant in the fit can be tuned in such a way that the $\chi^2$ of the pertinent fit can have 
a relatively low value. However, these fits do not have an isospin zero pole on the unphysical
sheet at a position which could be associated with the $\Lambda(1405)$.
In fact, if the $\pi \Sigma$ event 
distribution were simply approximated by the invariant mass distribution of $I=0$ $\pi \Sigma$
states, see e.g.\ \cite{KSW1, OR}, one would 
observe no peak structure at all for these fits.
Taking the well-established four star resonance $\Lambda(1405)$ for granted, one
should identify at least one pole of the $T$-matrix in the near vicinity,
and fits without a nearby pole must be dropped.

Finally, there is no experimental indication for an $S=-1$, $I=1$ $s$-wave baryon 
resonance below $\bar{K} N$ threshold, the lowest possible candidates being $\Sigma(1480)$
(one star resonance) and $\Sigma(1560)$ (two star resonance) which are listed as ``bumps'' 
in \cite{pdg}. While spin and parity of both states have not been determined yet, 
recent experiments \cite{CrBa2, ANKE} yield controversial results even on the existence of 
$\Sigma(1480)$. If, however, these low lying $I=1$ resonances 
should be confirmed in the future and have the required quantum 
numbers, their position is still above the relevant energy region considered here.
Therefore we drop fits which entail a pronounced $I=1$ resonance structure below $W = 1.44\GeV$
caused by an isospin one pole close and immediately connected to the real axis. 
More precisely, fits with $I=1$ poles located at $\Im(W) > -50\MeV$, i.e. less than 
twice as far from the real axis as typical $\Lambda(1405)$ poles, are not taken into account.

We observe fits which agree 
with the DEAR results at a lower overall $\chi^2/\mbox{d.o.f.}$ than indicated in 
Fig.~\ref{fig:KHyd} (but still outside the 1$\sigma$ confidence region)
if the above mentioned criteria are omitted, 
e.g., a $\chi^2/\mbox{d.o.f.} = 2.0$ value is obtained in the full approach.
All of these fits have an isospin one pole in common which is very close to the $\bar{K}N$ thresholds 
and either a few
MeV above (i.e.\ on the physical sheet) or below (i.e.\ on an unphysical sheet) the real axis.
Solutions of this type have been reported on in \cite{OPV, Oller}. However, such fits clearly 
violate one of the criteria discussed above and are not considered here.

In the remainder of this section we will focus on the resonance pole structure of the $\Lambda(1405)$,
i.e., the $I=0$ poles that are located on the unphysical sheet which is directly connected 
to the physical real axis between the $\pi \Sigma$ and $\bar{K} N$ thresholds.
The nature of the $\Lambda(1405)$ has recently attracted considerable
interest. It has been claimed that instead of the usual appearance of one resonance pole
the $\Lambda(1405)$ results from a pronounced two-pole structure with both poles
being very close to the physical region \cite{JOORM}. While only the
Weinberg-Tomozawa contact interaction was taken into account in \cite{JOORM},
the inclusion of the next-to-leading order contact terms
destroyed the pronounced two-pole structure, as one pole was shifted further away 
from the real axis and its contribution to the physical region dissolved in the background
\cite{BNW1}. As already pointed out in \cite{BNW1}, the position of this pole depends very sensitively 
on the values of the parameters, whereas the other one remains relatively fixed and close to the 
physical axis. In \cite{OPV}, e.g., which also includes the next-to-leading order contact interactions
the pole is located at $1321 - i\,43.5\MeV$, i.e., even {\it below} the $\pi \Sigma$ threshold(s)
and hence not immediately connected to the physical region, whereas 
in the ``WT'' approach a pole around a mass of 1390 MeV was found in \cite{JOORM}.

For most fits in the 1$\sigma$ confidence interval we observe two isospin zero poles in the 
region $\Re{W} = (1250 \ldots 1600)\MeV$, $\Im{W} = (-2.5 \ldots\ $$-250)\MeV$. 
In some cases, however, it happens that there is only one $I=0$ pole in this region. We
then extend the pole search beyond the chosen limits until a second $I=0$ pole is found.
The observed pole positions are compiled in Table~\ref{tab:polepos}, where ``first pole''
refers to the pole which is closer to the real axis at $1.405\GeV$, i.e., the position 
of the $\Lambda(1405)$ peak. While the variation of the position of this first pole is 
remarkably small in all three approaches, ``WT'', ``WTB'', ``full'', the position of the 
second $I=0$ pole scatters over a wide range in the complex $W$ plane and consequently does 
not have in all fits a significant impact on physical observables.

\begin{table}[t]
\centering
\begin{tabular}{|c|r|r|}
\hline
& \multicolumn{2}{c|}{pole position (MeV)} \\
& \multicolumn{1}{c|}{first pole} & \multicolumn{1}{c|}{second pole} \\
\hline
``WT''   & $1420^{+19}_{-16}  - i\,\bigl(20^{-8}_{+22}\bigr)$
         & $1440^{+56}_{-227}  - i\,\bigl(76^{-51}_{+51}\bigr)$ \\
\hline
``WTB''  & $1423^{+10}_{-12}  - i\,\bigl(15^{-8}_{+20}\bigr)$
         & $1366^{+122}_{-20} - i\,\bigl(84^{-37}_{+121}\bigr)$ \\
\hline
``full'' & $1418^{+60}_{-38}  - i\,\bigl(31^{-24}_{+34}\bigr)$
         & $1348^{+293}_{-86} - i\,\bigl(62^{-58}_{+212}\bigr)$ \\
\hline
\end{tabular}
\caption{Positions of the first and second $I=0$ pole in the complex $W$ plane.
         The central values correspond to the best fit in each approach, while the error 
         ranges include all fits in the $1\sigma$ confidence region.}
\label{tab:polepos}
\end{table}

We also observe very few fits with a third $I=0$ pole which appears either at 
$\Re{W} > 1.5\GeV$ and thus well above the $\bar{K} N$ threshold(s) (recall that the pertinent 
Riemann sheet is connected to the real axis below these thresholds) or deep in the complex $W$ plane
($\Im{W} < -150\MeV$). In these fits, the third pole is thus not expected to have much 
influence on physical observables.

From the discussions above it becomes clear that the analytic continuation
to the complex energy plane and the resulting pole positions depend sensitively 
on the dynamical input of the chiral SU(3) effective Lagrangian.
A rigorous extraction of the pole positions, in particular the second one,
appears therefore very unlikely from the experimental data considered in the present 
investigation. Additional experimental input, however, may help to further constrain 
the position of the second pole, see e.g. Ref.~\cite{MOR}.
But in any case, as illustrated above, the pole positions can very well
serve as an additional constraint to rule out some fits.

\section{Conclusions}\label{sec:summary}

In this work, we have provided an extraction of the strong $K^- p$ scattering length
derived within chiral unitary approaches from fits to available low-energy
$K^- p$ scattering data. To this end, we have utilized three variants of such
chiral unitary approaches which differ in the choice of the interaction kernel.
In the first approach, the interaction kernel is derived from the Weinberg-Tomozawa
contact interaction at leading chiral order which is successively supplemented
by the Born terms and the contact interactions of next-to-leading chiral order
in the second and third framework, respectively. The usage of three different
interaction kernels helps to estimate the inherent model dependence of such approaches.

For all three approaches a least-squares fit to low-energy data in $S=-1$ meson-baryon
channels is performed. These are in detail $K^-p$ scattering, the $\pi \Sigma$
mass spectrum and the precisely measured $K^-p$ threshold decay ratios.
Fits with the lowest $\chi^2/\mbox{d.o.f.}$ value are found in the full approach
including the higher order couplings, while the $\chi^2/\mbox{d.o.f.}$ value is
largest in the Weinberg-Tomozawa approach.

Based on a very large variety of different fits to data we can provide an error range
for the strong $K^- p$ scattering length 
which is related to the strong interaction shift and width in kaonic hydrogen.
We obtain an energy shift and width in kaonic hydrogen which is in agreement
with the KEK experiment, but disagrees with DEAR. The present analysis confirms
the findings of \cite{BNW1} by pointing on questions of consistency of the
recent DEAR measurement with previous $K^- p$ scattering data. 
The conservative
error range for $a_{K^- p}$ derived from chiral unitary approaches is in clear
disagreement with the one deduced from the DEAR experiment.

Furthermore, we have critically investigated the pole structure of the fits.
The first isospin zero pole remains relatively fixed in all fits and close
to the physical axis, whereas the second pole is quite 
sensitive to the chosen parameters of the approach. In particular, the influence of the second pole 
on physical observables is substantially reduced, if it is further away from the real
axis, and can even dissolve in the background. 
Although the pole positions depend sensitively 
on the dynamical input of the chiral SU(3) effective Lagrangian,
we have illustrated that the general pole structure of a fit
can serve as an additional criterion to consider the fit as unphysical.
In this respect, we look very much  forward to the electromagnetic production 
data of the $\Lambda (1405)$ from the ELSA accelerator at Bonn
which may help to further clarify the pole 
structure of the $K^- p$ scattering amplitude below threshold.

\section*{Acknowledgments}\label{sec:Ackno}

We thank U.~Raha and A.~Rusetsky for useful discussions. Special thanks go to M.~Belushkin for 
his valuable help and assistance with the implementation of the minimization routines.
Partial financial support by Deutsche Forschungsgemeinschaft 
(SFB/TR 16, ``Subnuclear Structure of Matter'', and BO 1481/6-1)
is gratefully acknowledged.
This research is part of the EU Integrated Infrastructure Initiative Hadron Physics under contract
number RII3-CT-2004-506078.

\appendix

\section{Results of the fit to scattering data}\label{app:plots}

\begin{figure}
\centering
\begin{tabular}{p{0.5cm}rp{1.0cm}r}
 &
\begin{overpic}[height=0.20\textheight,clip]{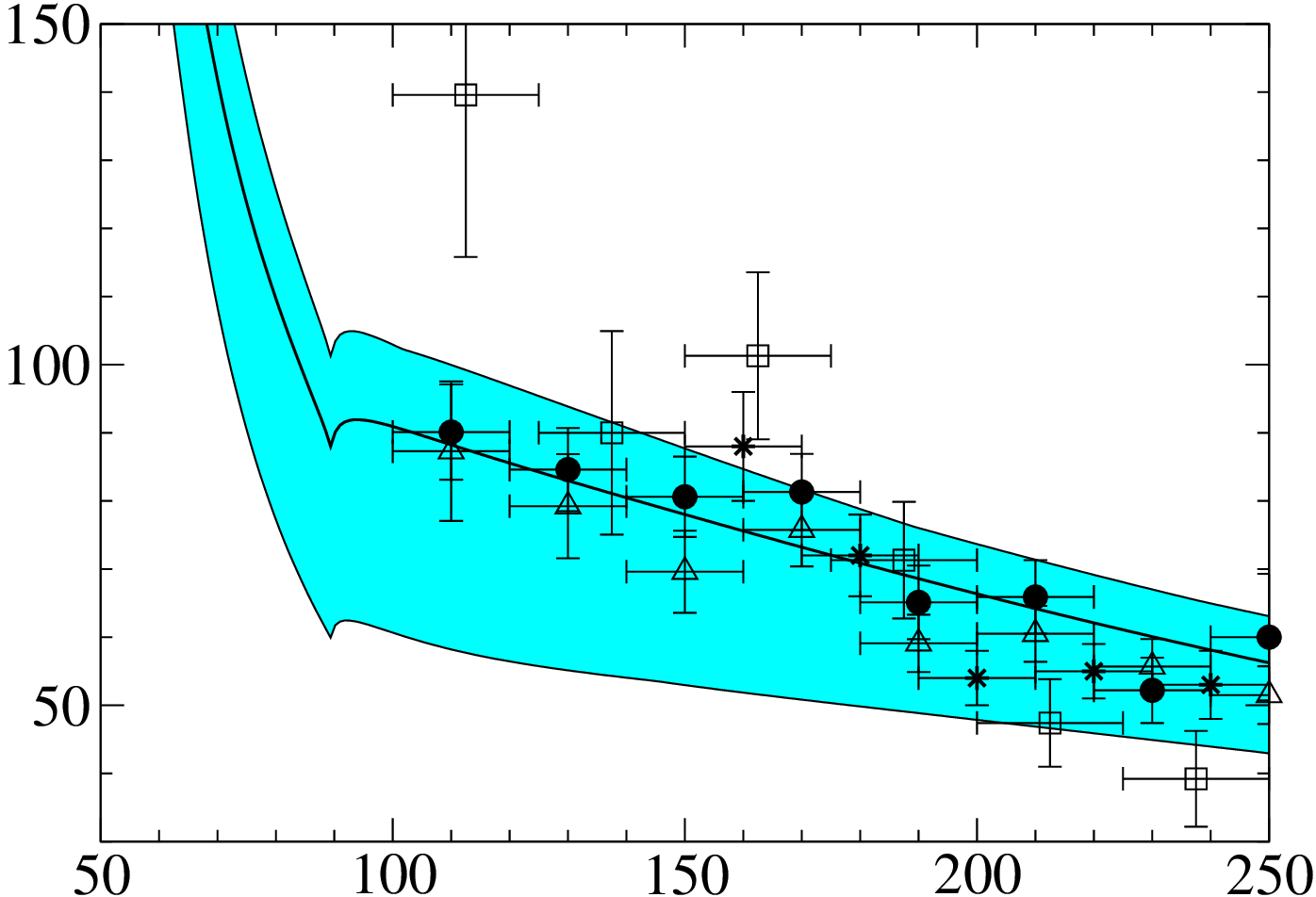}
  \put(-12,9){\rotatebox{90}{{\scalebox{0.9}{$\sigma [K^- p \to K^- p]$ (mb)}}}}
\end{overpic} & &
\begin{overpic}[height=0.20\textheight,clip]{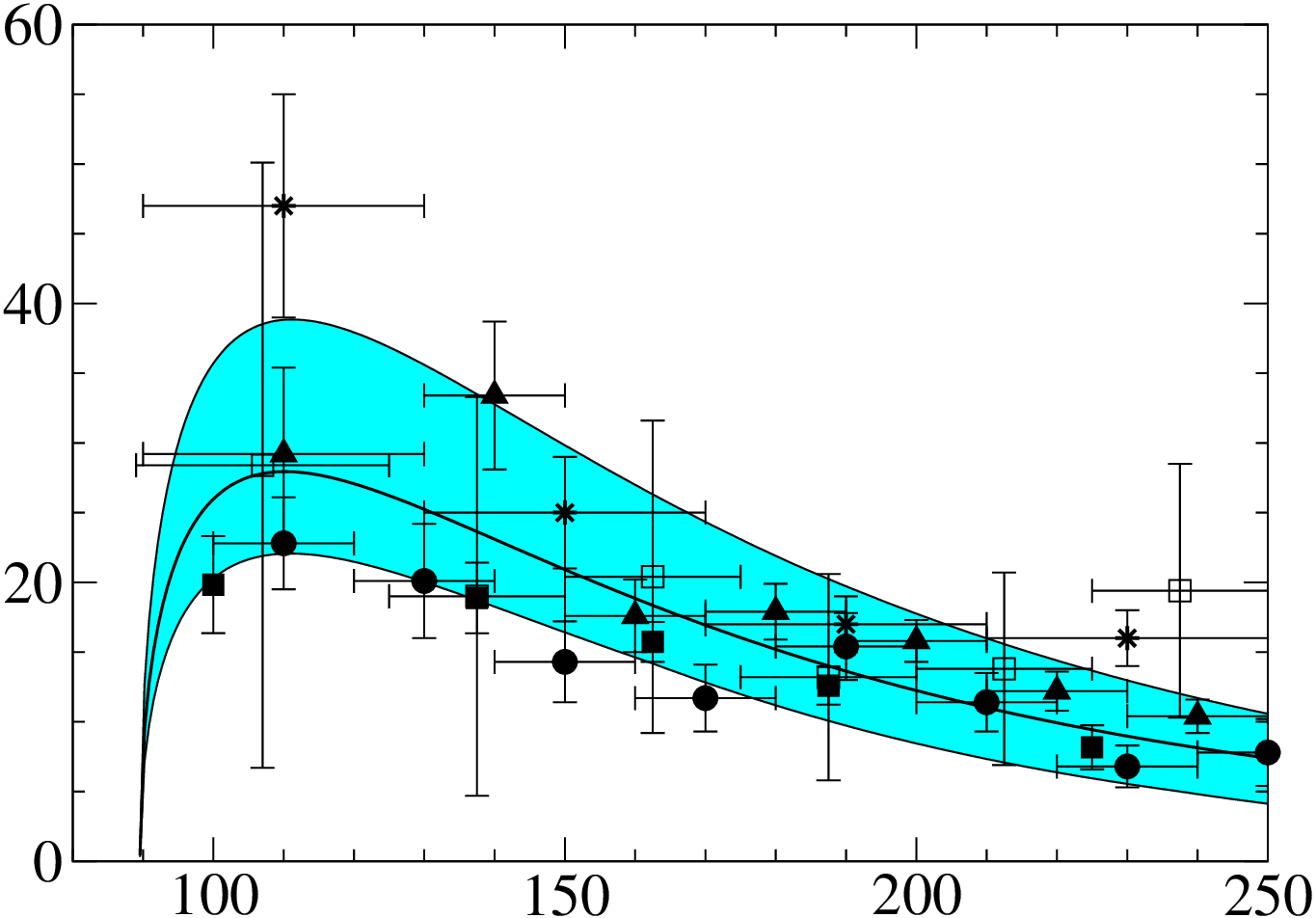}
  \put(-12,9){\rotatebox{90}{{\scalebox{0.9}{$\sigma [K^- p \to \bar{K}^0 n]$ (mb)}}}}
\end{overpic} \\[0.02\textheight]
 &
\begin{overpic}[height=0.20\textheight,clip]{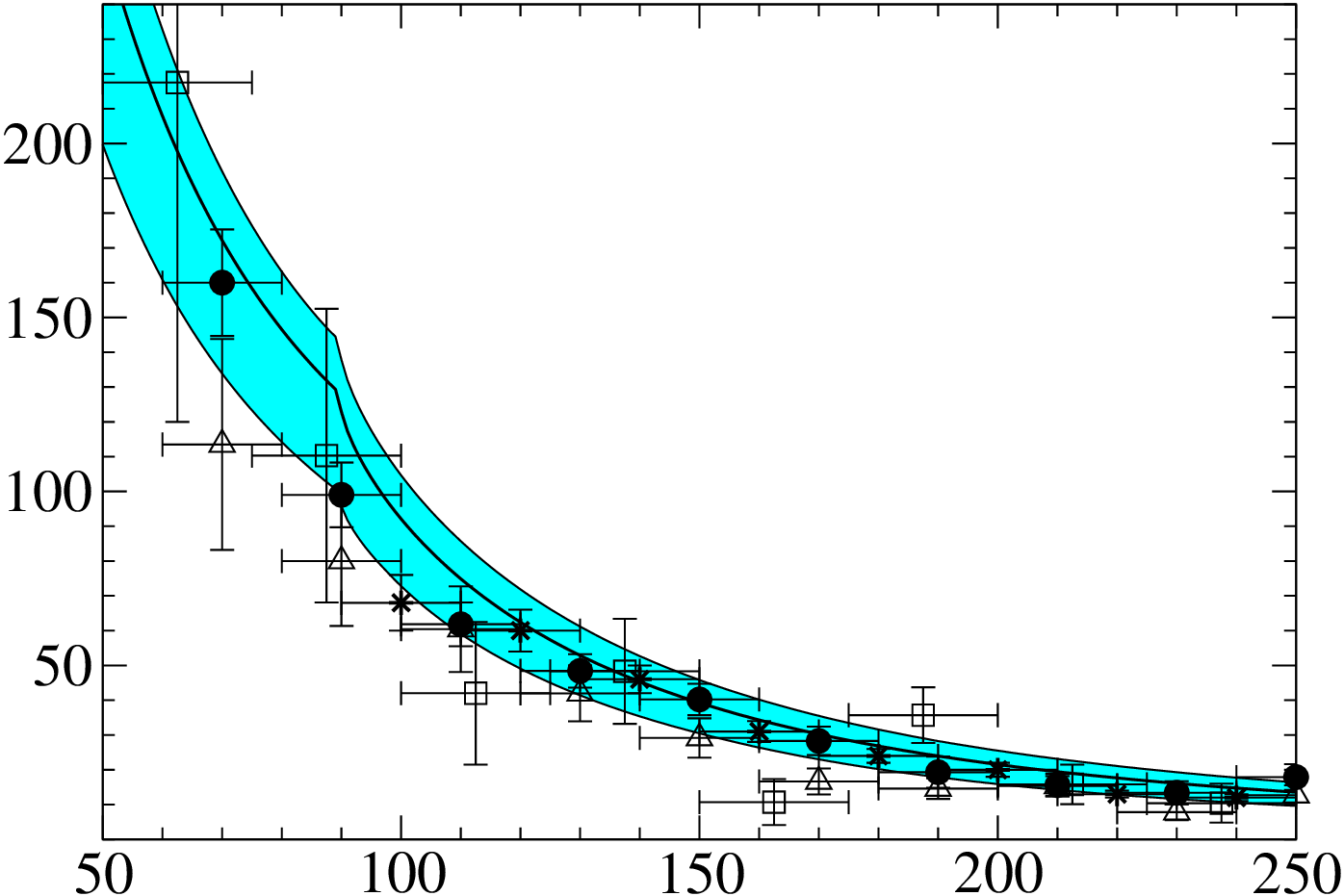}
  \put(-10,9){\rotatebox{90}{{\scalebox{0.9}{$\sigma [K^- p \to \pi^+ \Sigma^-]$ (mb)}}}}
\end{overpic} & &
\begin{overpic}[height=0.20\textheight,clip]{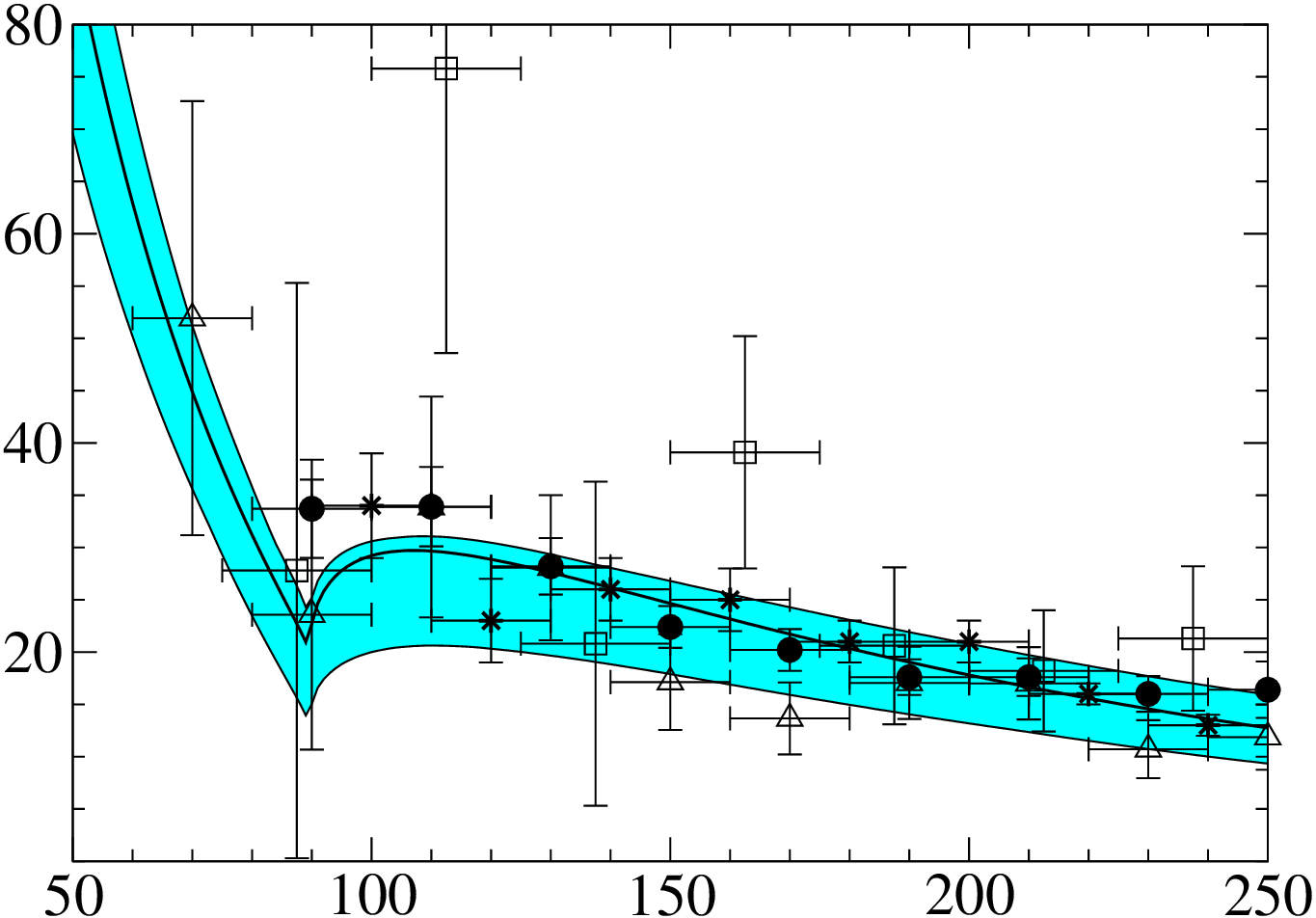}
  \put(-12,9){\rotatebox{90}{{\scalebox{0.9}{$\sigma [K^- p \to \pi^- \Sigma^+]$ (mb)}}}}
\end{overpic} \\[0.02\textheight]
 &
\begin{overpic}[height=0.20\textheight,clip]{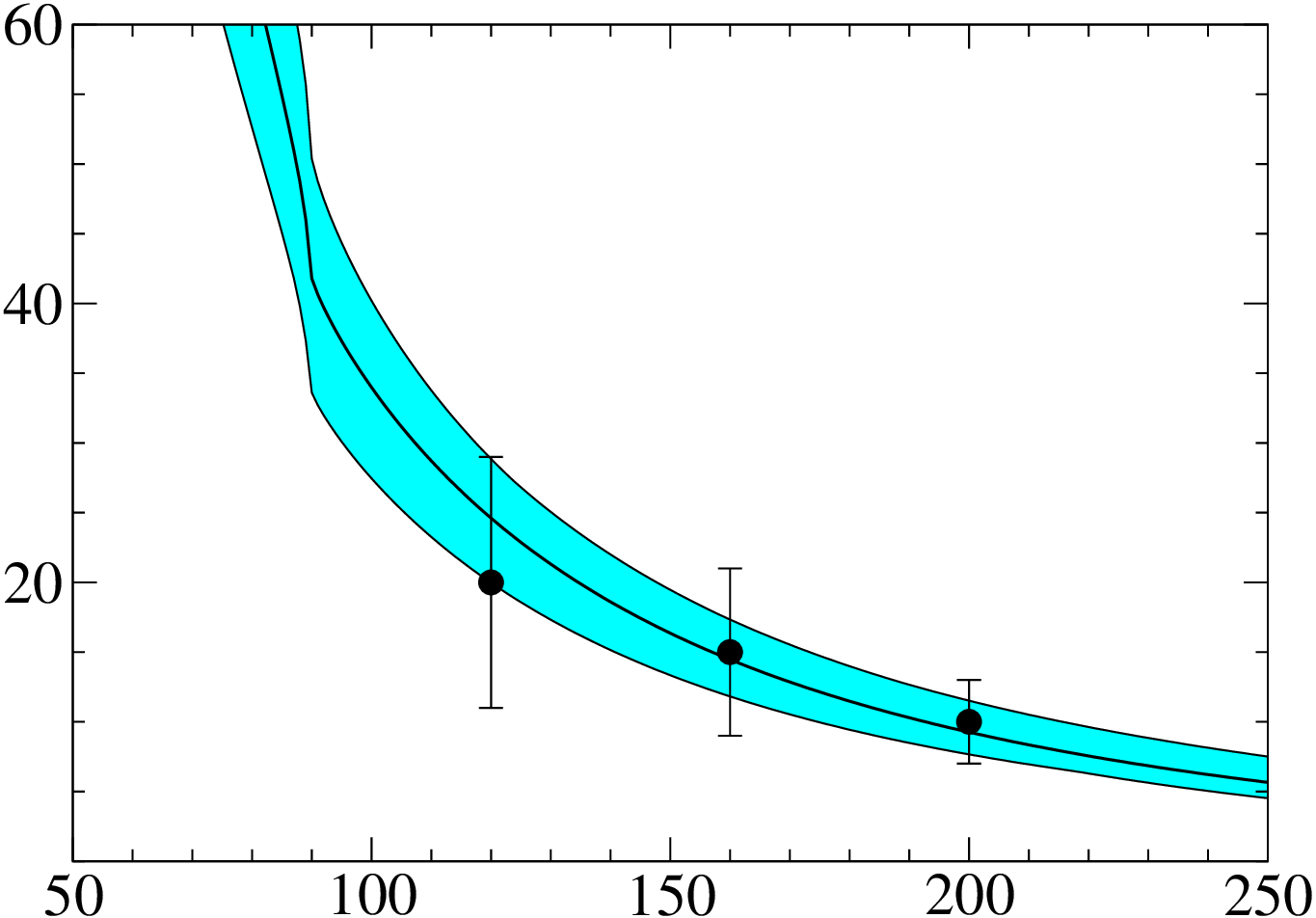}
  \put(-15,9){\rotatebox{90}{{\scalebox{0.9}{$\sigma [K^- p \to \pi^0 \Sigma^0]$ (mb)}}}}
  \put(5,-9){\scalebox{0.9}{incident $K^-$ lab.\  momentum (MeV/c)}}
\end{overpic} & &
\begin{overpic}[height=0.20\textheight,clip]{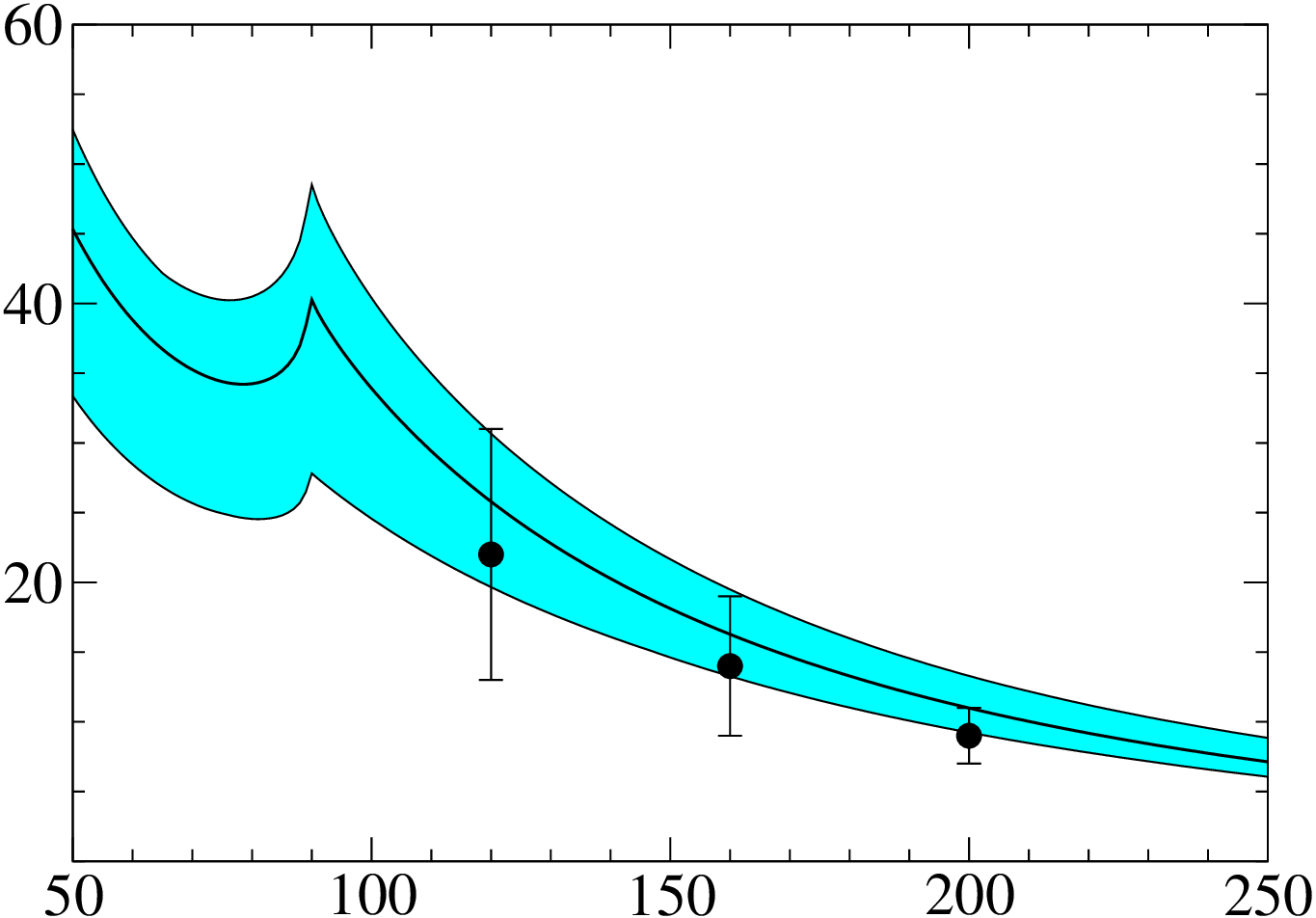}
  \put(-12,7){\rotatebox{90}{{\scalebox{0.9}{$\sigma [K^- p \to \pi^0 \Lambda]$ (mb)}}}}
  \put(5,-9){\scalebox{0.9}{incident $K^-$ lab.\  momentum (MeV/c)}}
\end{overpic}
\end{tabular}
\vspace{2.0ex}
\caption{Total cross sections for $K^- p$ scattering into various channels calculated in the
         ``WT'' approach. The best fit is represented by the solid line while the shaded 
         area indicates the 1$\sigma$ confidence region. The data
         are taken from  \cite{Hum} (empty squares), \cite{Sak} (empty triangles),
         \cite{Kim} (filled circles), \cite{Kit} (filled squares), 
         \cite{Eva} (filled triangles), \cite{Cib} (stars).}
\label{fig:CSWT}
\end{figure}

\begin{figure}
\centering
\begin{tabular}{p{0.5cm}rp{1.0cm}r}
 &
\begin{overpic}[height=0.20\textheight,clip]{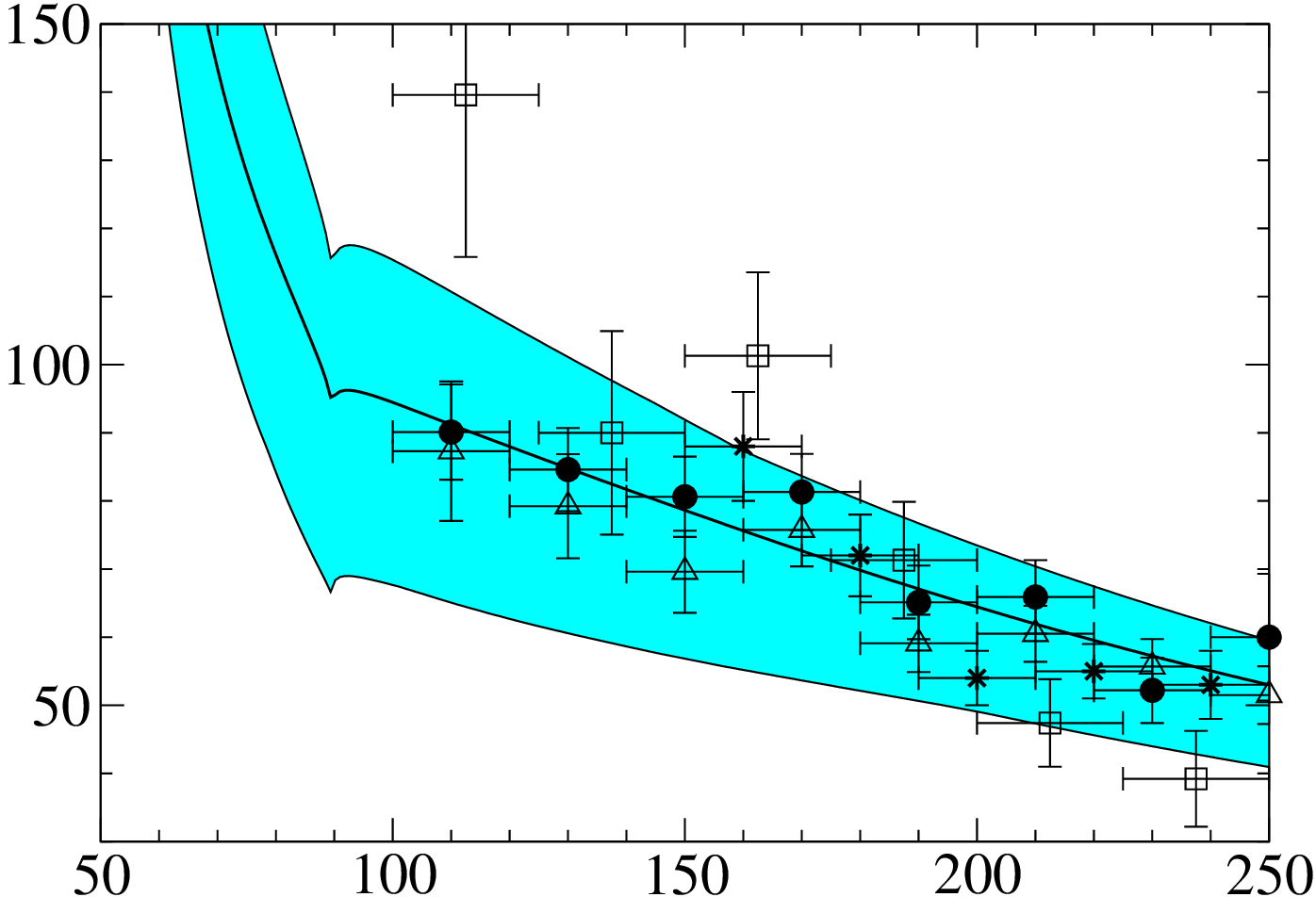}
  \put(-12,9){\rotatebox{90}{{\scalebox{0.9}{$\sigma [K^- p \to K^- p]$ (mb)}}}}
\end{overpic} & &
\begin{overpic}[height=0.20\textheight,clip]{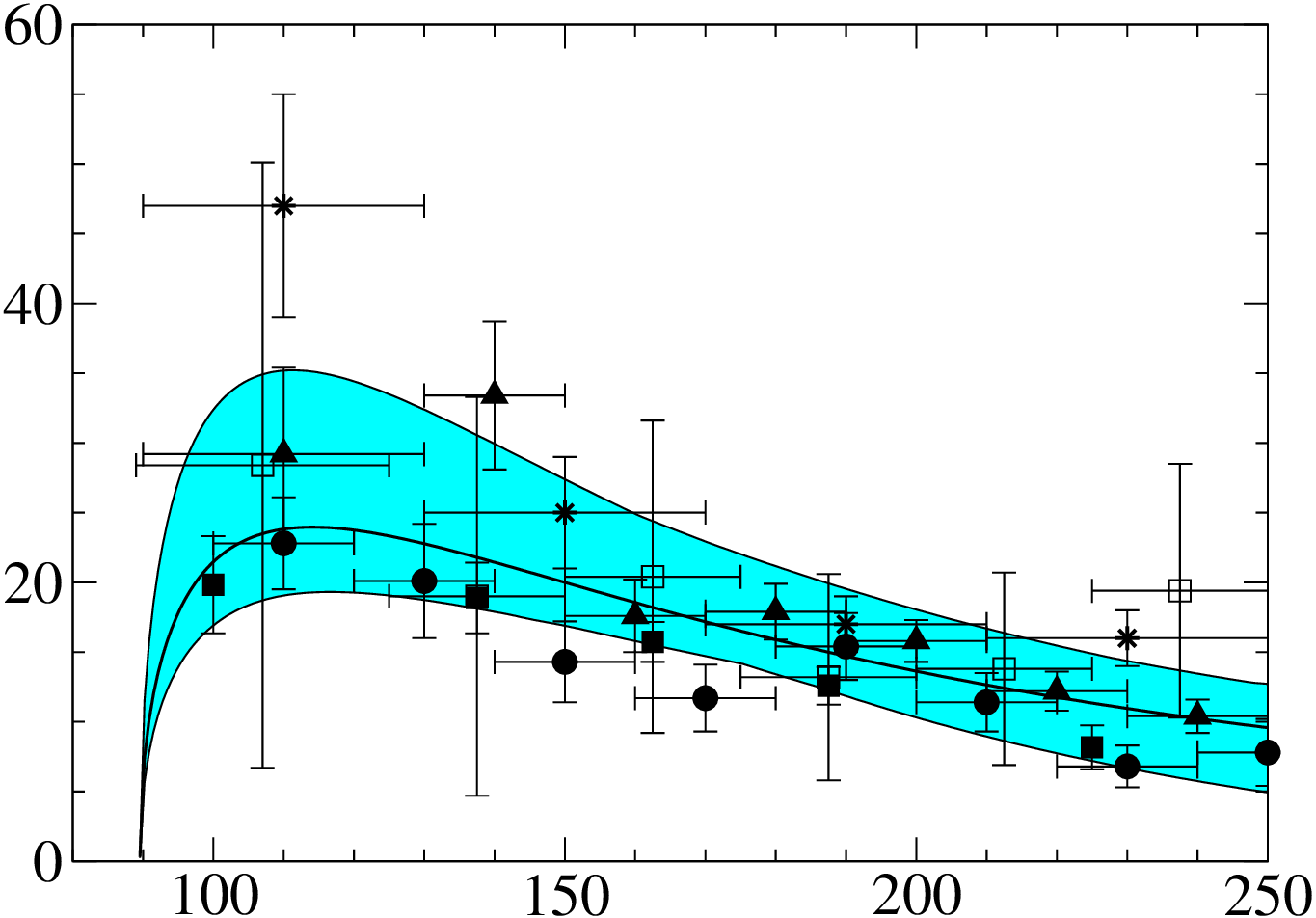}
  \put(-12,9){\rotatebox{90}{{\scalebox{0.9}{$\sigma [K^- p \to \bar{K}^0 n]$ (mb)}}}}
\end{overpic} \\[0.02\textheight]
 &
\begin{overpic}[height=0.20\textheight,clip]{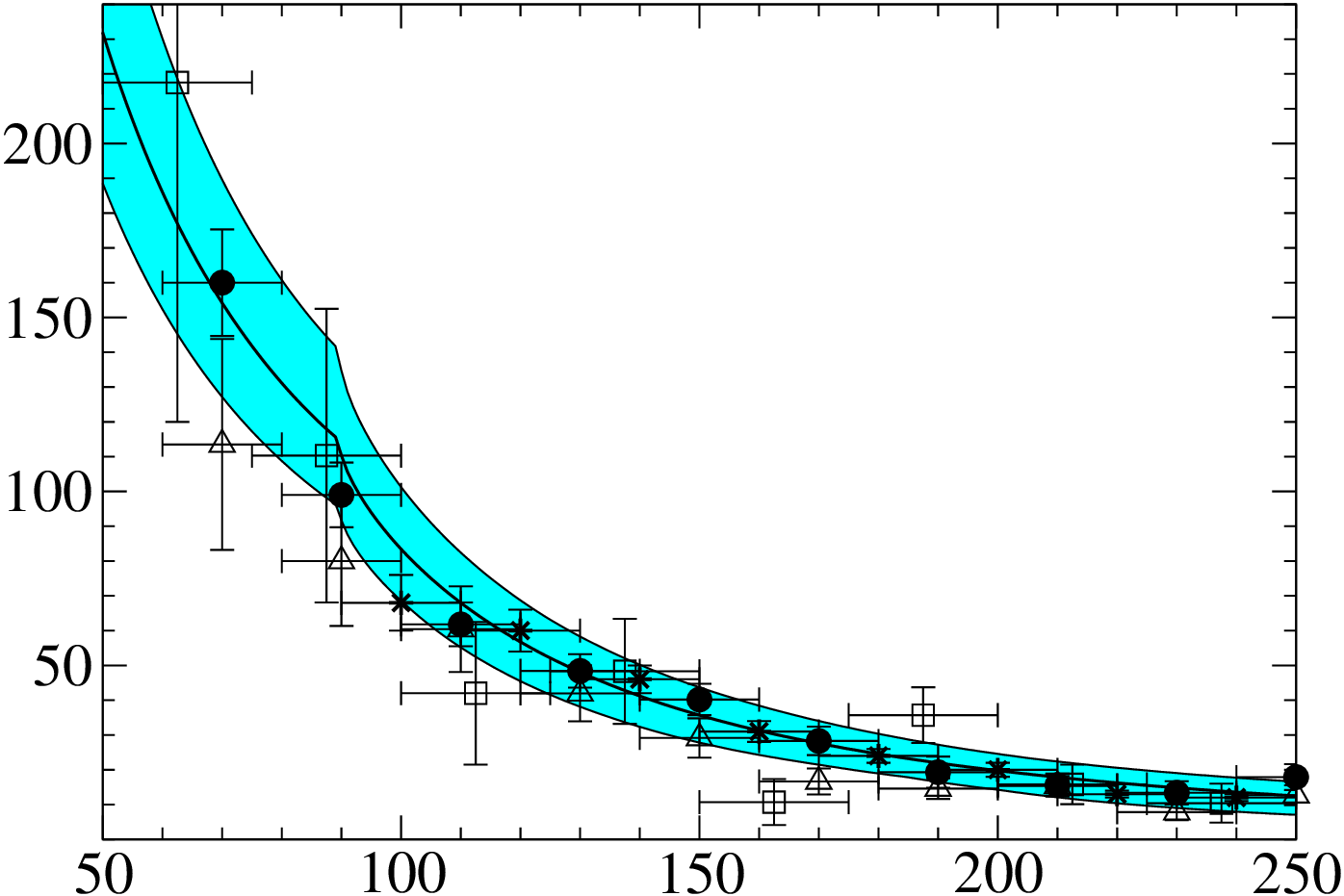}
  \put(-10,9){\rotatebox{90}{{\scalebox{0.9}{$\sigma [K^- p \to \pi^+ \Sigma^-]$ (mb)}}}}
\end{overpic} & &
\begin{overpic}[height=0.20\textheight,clip]{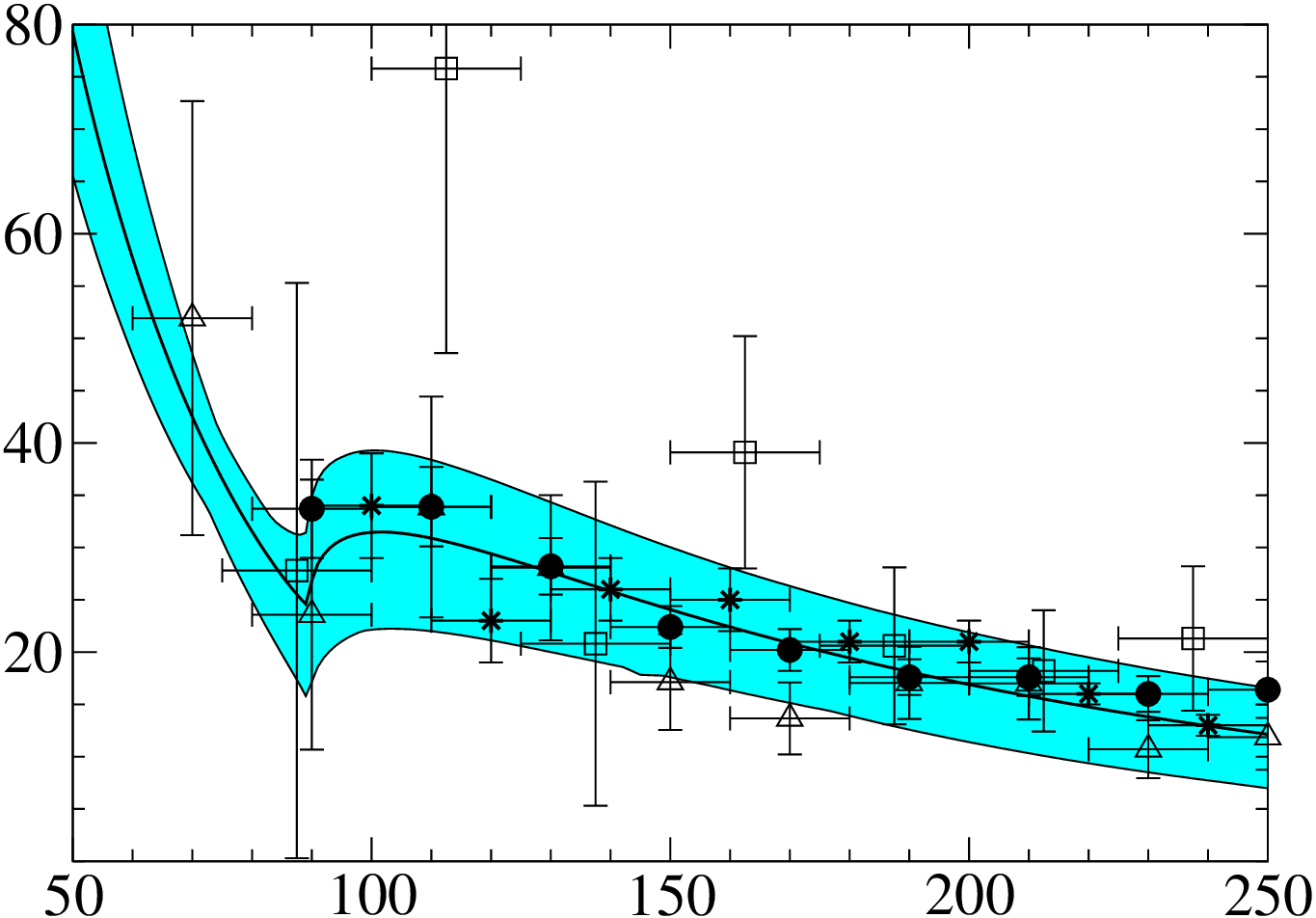}
  \put(-12,9){\rotatebox{90}{{\scalebox{0.9}{$\sigma [K^- p \to \pi^- \Sigma^+]$ (mb)}}}}
\end{overpic} \\[0.02\textheight]
 &
\begin{overpic}[height=0.20\textheight,clip]{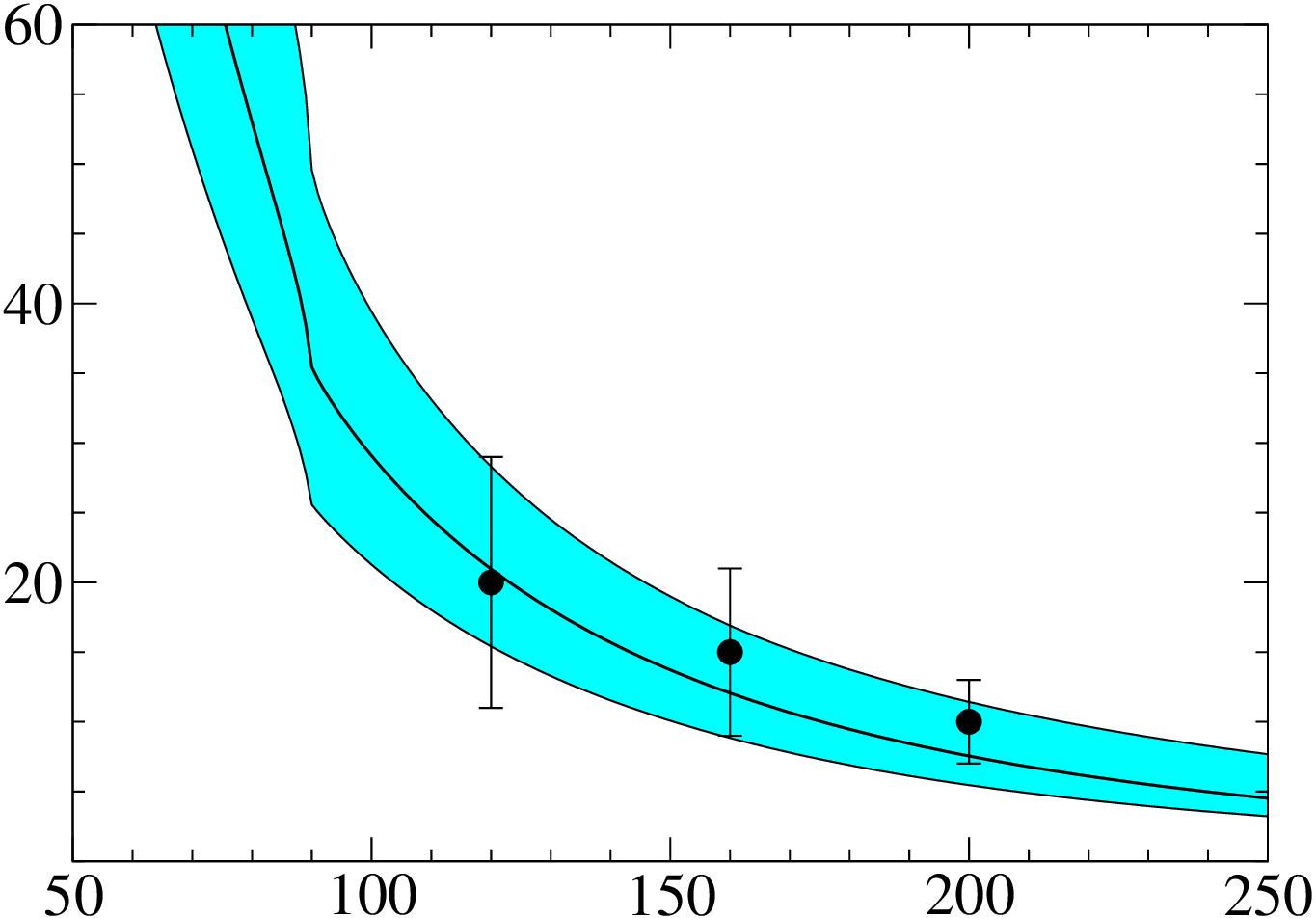}
  \put(-15,9){\rotatebox{90}{{\scalebox{0.9}{$\sigma [K^- p \to \pi^0 \Sigma^0]$ (mb)}}}}
  \put(5,-9){\scalebox{0.9}{incident $K^-$ lab.\  momentum (MeV/c)}}
\end{overpic} & &
\begin{overpic}[height=0.20\textheight,clip]{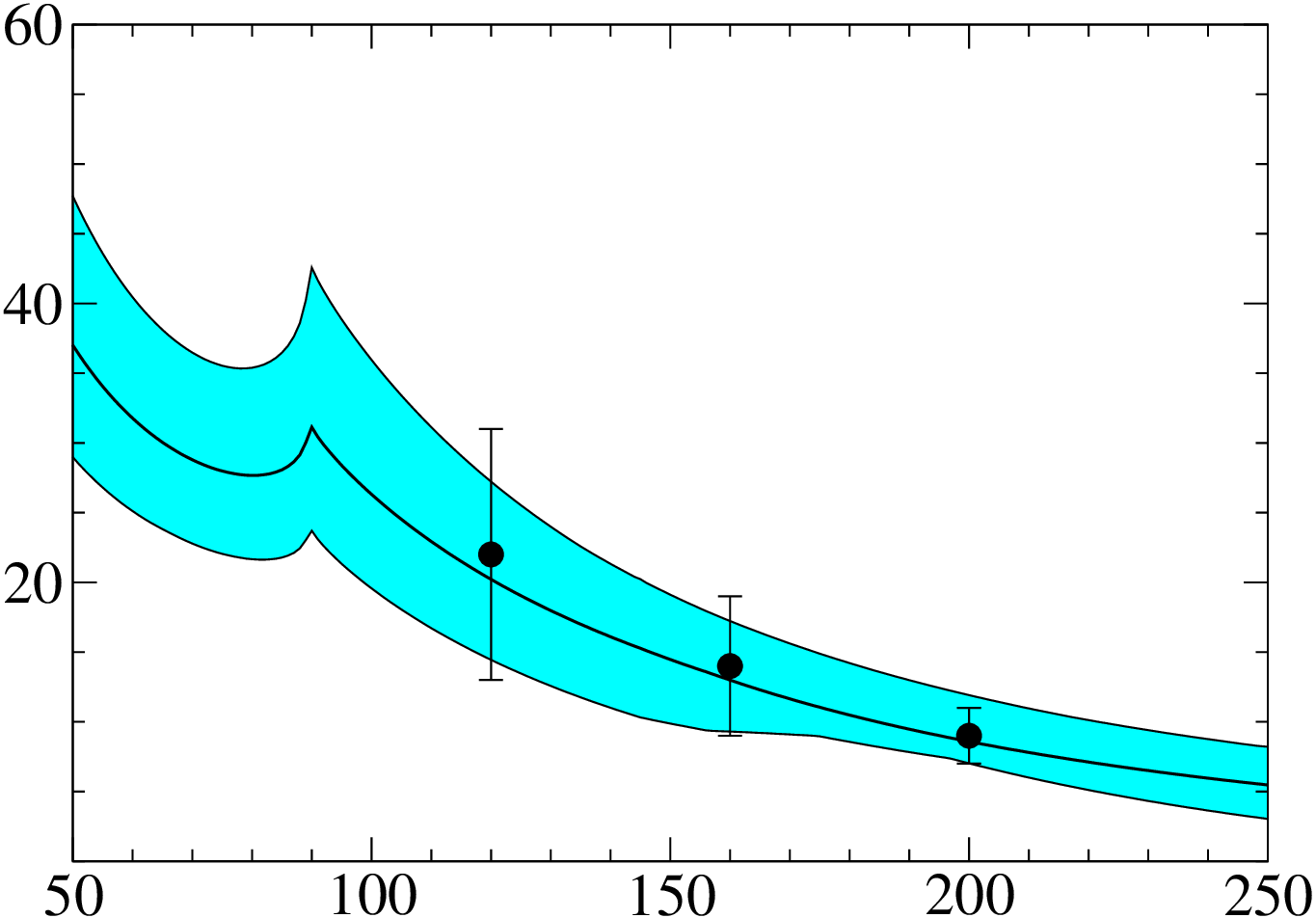}
  \put(-12,7){\rotatebox{90}{{\scalebox{0.9}{$\sigma [K^- p \to \pi^0 \Lambda]$ (mb)}}}}
  \put(5,-9){\scalebox{0.9}{incident $K^-$ lab.\  momentum (MeV/c)}}
\end{overpic}
\end{tabular}
\vspace{2.0ex}
\caption{Total cross sections for $K^- p$ scattering into various channels calculated in the
         ``WTB'' approach. The best fit is represented by the solid line while the shaded 
         area indicates the 1$\sigma$ confidence region. The data
         are taken from  \cite{Hum} (empty squares), \cite{Sak} (empty triangles),
         \cite{Kim} (filled circles), \cite{Kit} (filled squares), 
         \cite{Eva} (filled triangles), \cite{Cib} (stars).}
\label{fig:CSC}
\end{figure}

\begin{figure}
\centering
\begin{tabular}{p{0.5cm}rp{1.0cm}r}
 &
\begin{overpic}[height=0.20\textheight,clip]{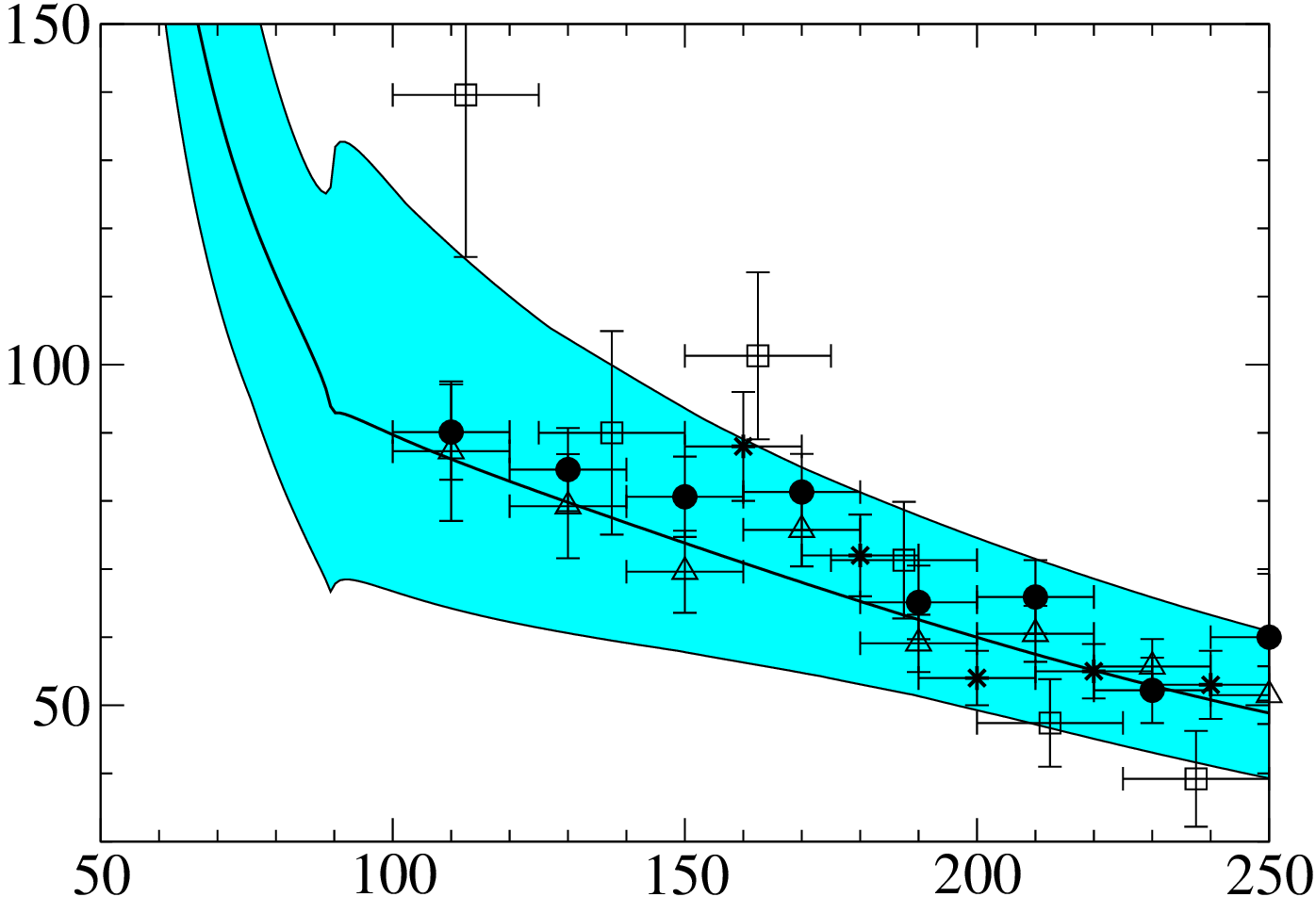}
  \put(-12,9){\rotatebox{90}{{\scalebox{0.9}{$\sigma [K^- p \to K^- p]$ (mb)}}}}
\end{overpic} & &
\begin{overpic}[height=0.20\textheight,clip]{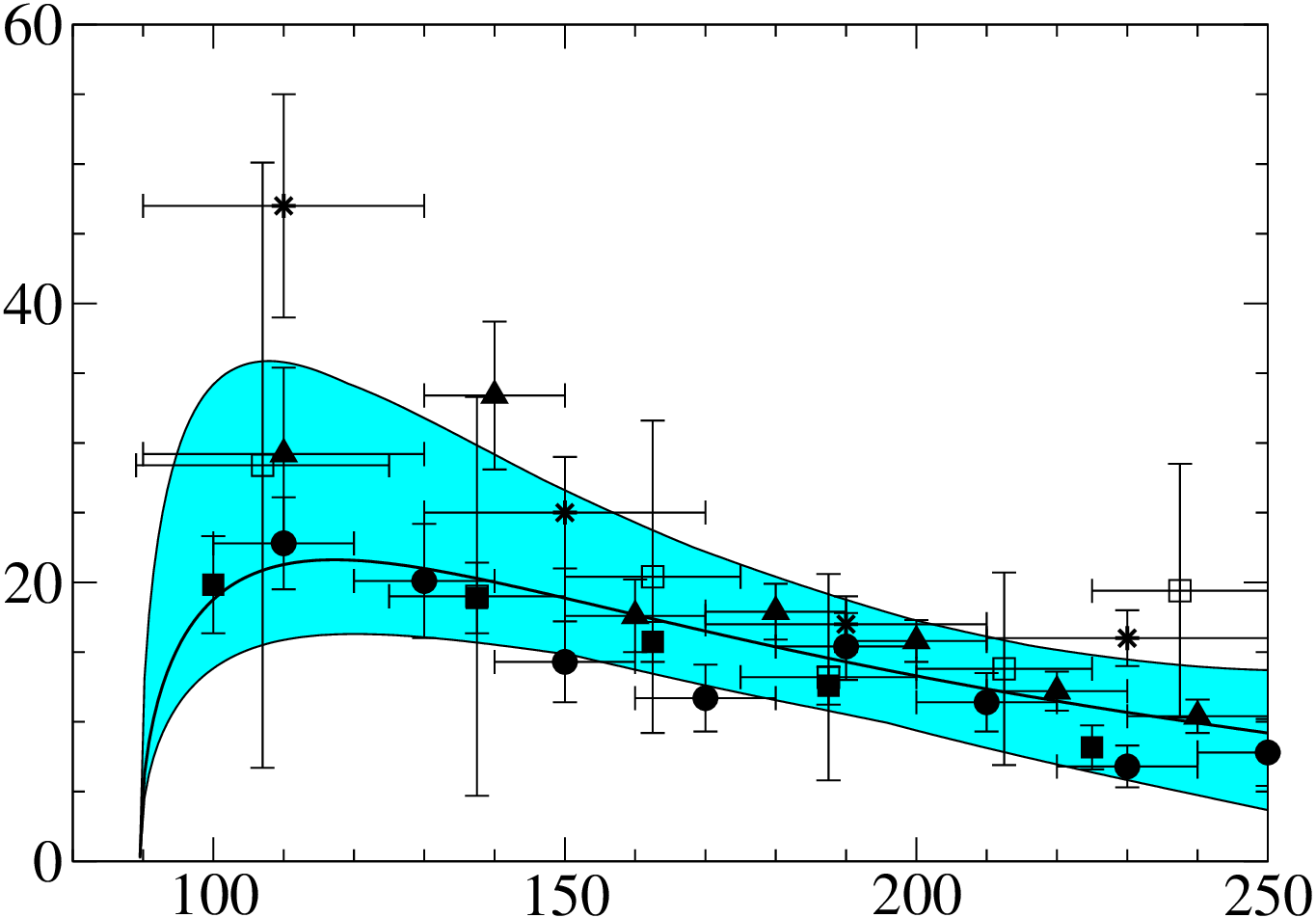}
  \put(-12,9){\rotatebox{90}{{\scalebox{0.9}{$\sigma [K^- p \to \bar{K}^0 n]$ (mb)}}}}
\end{overpic} \\[0.02\textheight]
 &
\begin{overpic}[height=0.20\textheight,clip]{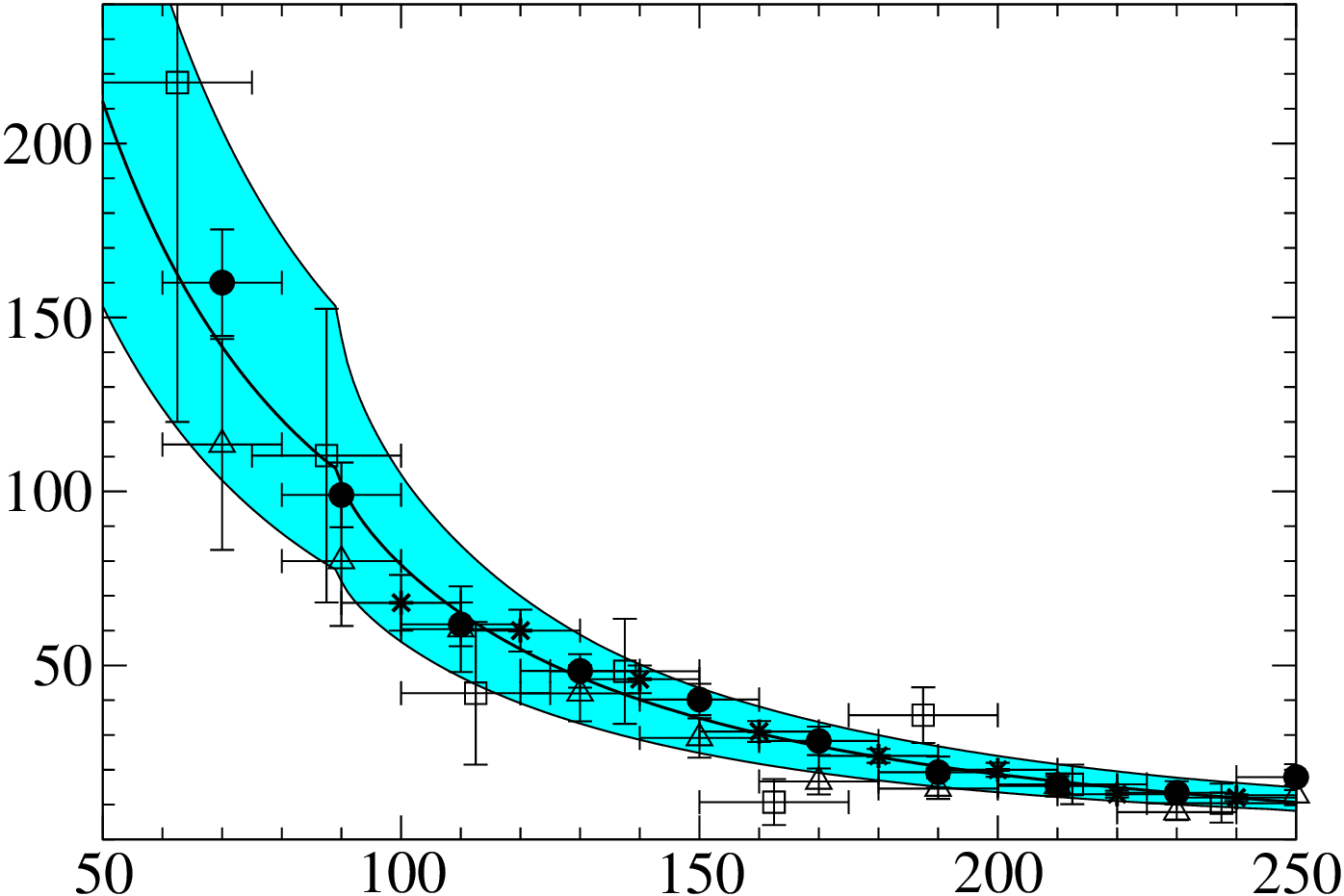}
  \put(-10,9){\rotatebox{90}{{\scalebox{0.9}{$\sigma [K^- p \to \pi^+ \Sigma^-]$ (mb)}}}}
\end{overpic} & &
\begin{overpic}[height=0.20\textheight,clip]{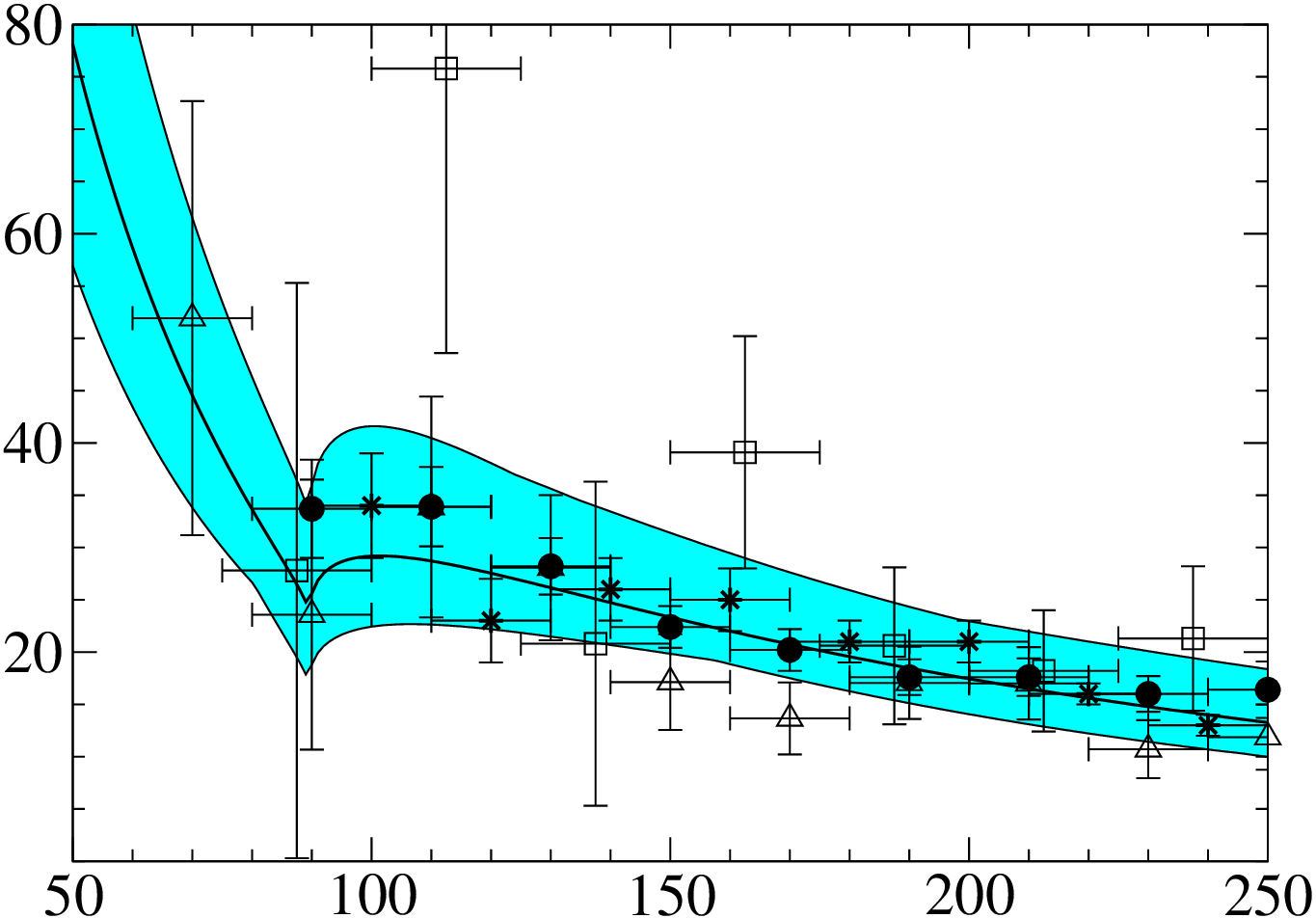}
  \put(-12,9){\rotatebox{90}{{\scalebox{0.9}{$\sigma [K^- p \to \pi^- \Sigma^+]$ (mb)}}}}
\end{overpic} \\[0.02\textheight]
 &
\begin{overpic}[height=0.20\textheight,clip]{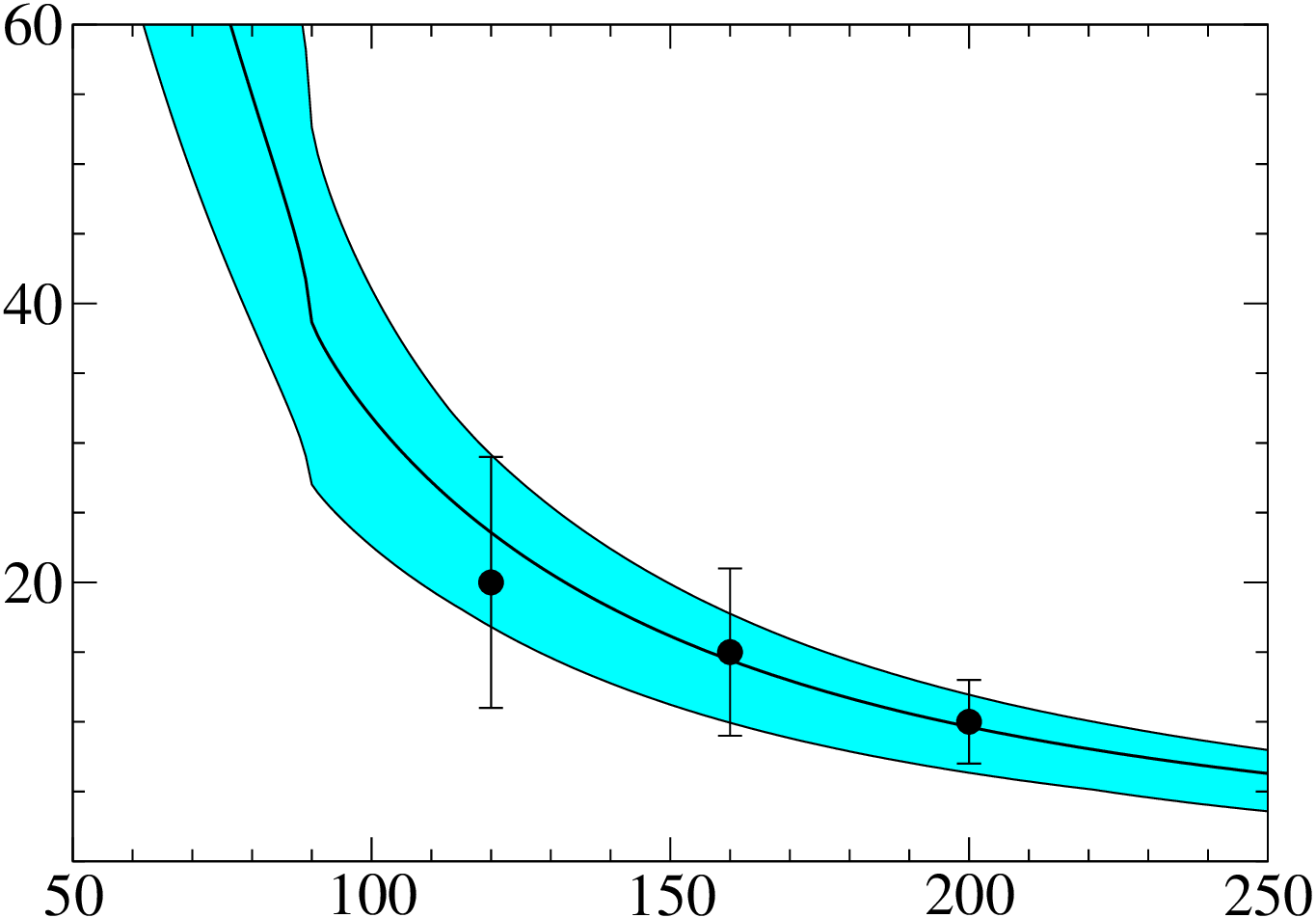}
  \put(-15,9){\rotatebox{90}{{\scalebox{0.9}{$\sigma [K^- p \to \pi^0 \Sigma^0]$ (mb)}}}}
  \put(5,-9){\scalebox{0.9}{incident $K^-$ lab.\  momentum (MeV/c)}}
\end{overpic} & &
\begin{overpic}[height=0.20\textheight,clip]{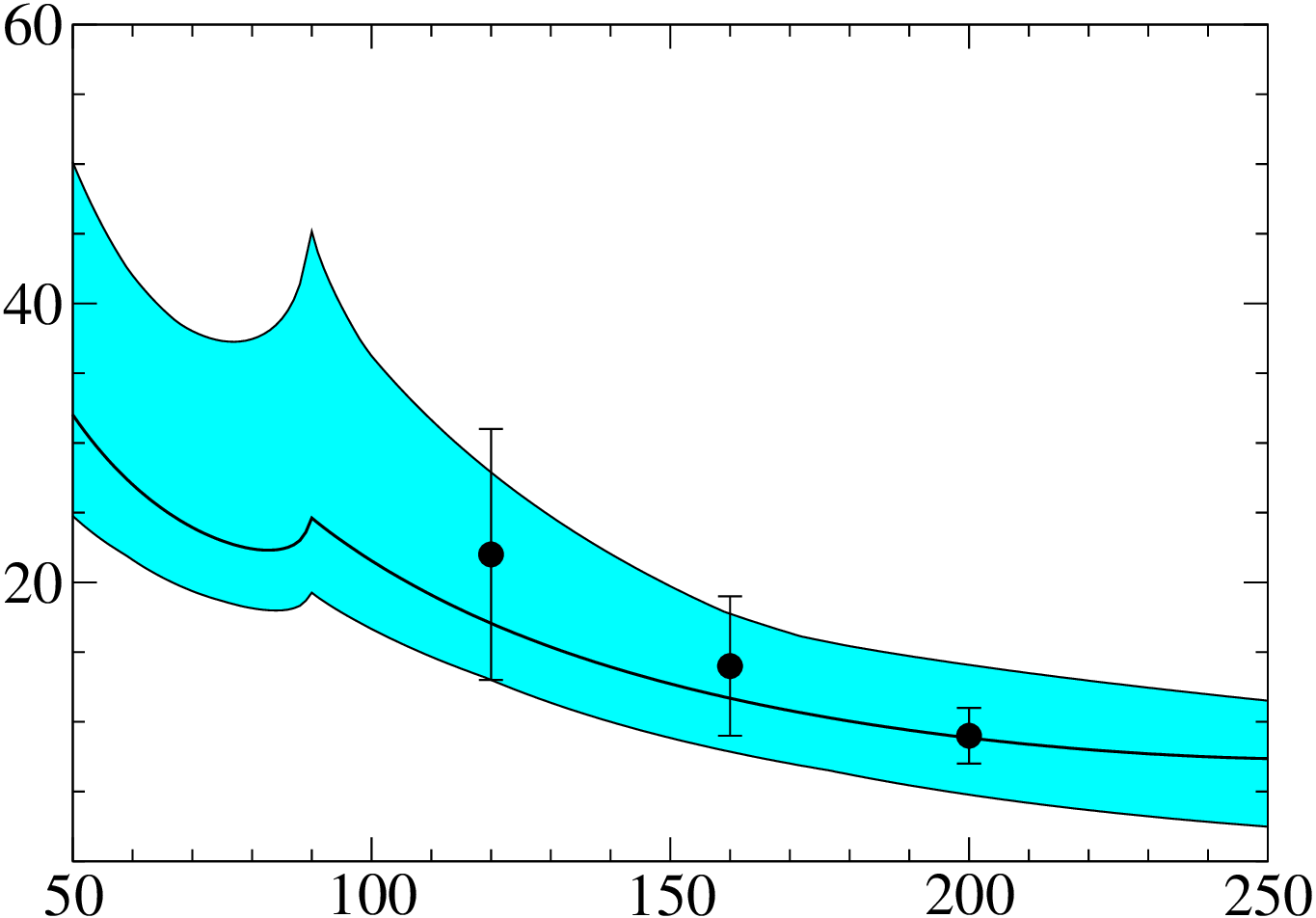}
  \put(-12,7){\rotatebox{90}{{\scalebox{0.9}{$\sigma [K^- p \to \pi^0 \Lambda]$ (mb)}}}}
  \put(5,-9){\scalebox{0.9}{incident $K^-$ lab.\  momentum (MeV/c)}}
\end{overpic}
\end{tabular}
\vspace{2.0ex}
\caption{Total cross sections for $K^- p$ scattering into various channels calculated in the
         ``full'' approach. The best fit is represented by the solid line while the shaded 
         area indicates the 1$\sigma$ confidence region. The data
         are taken from  \cite{Hum} (empty squares), \cite{Sak} (empty triangles),
         \cite{Kim} (filled circles), \cite{Kit} (filled squares), 
         \cite{Eva} (filled triangles), \cite{Cib} (stars).}
\label{fig:CSB}
\end{figure}

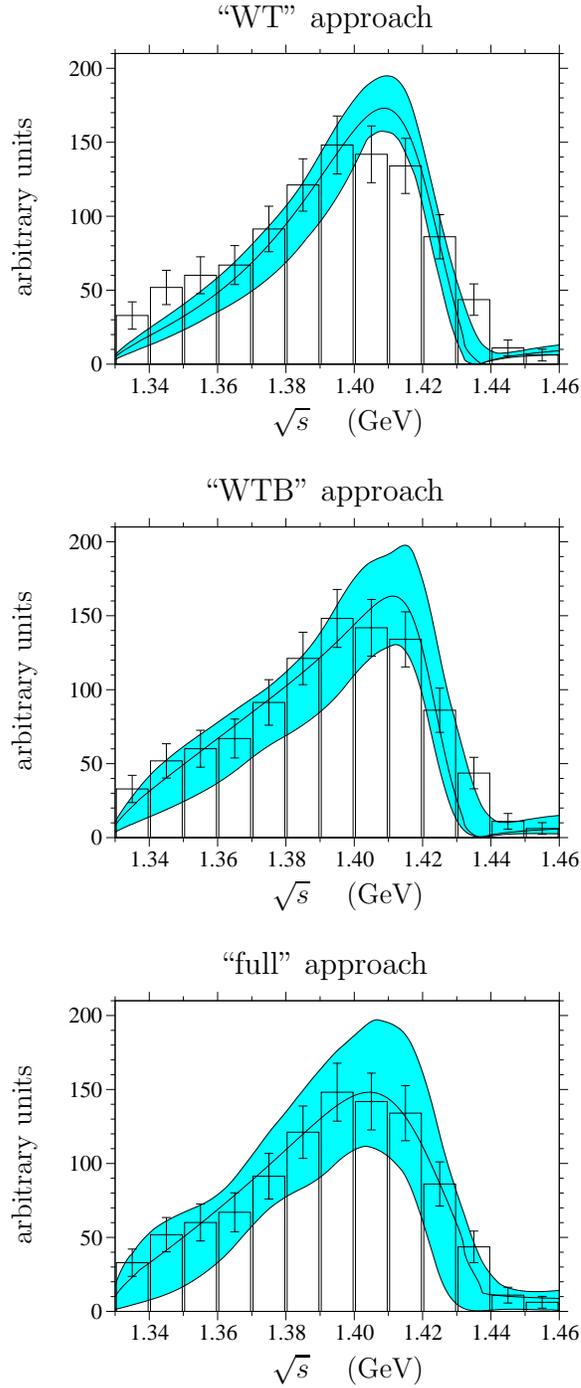
\begin{figure}
\centering
\begin{tabular}{c}
``WT'' approach \\[0.5ex]
\begin{overpic}[width=0.40\textwidth,clip]{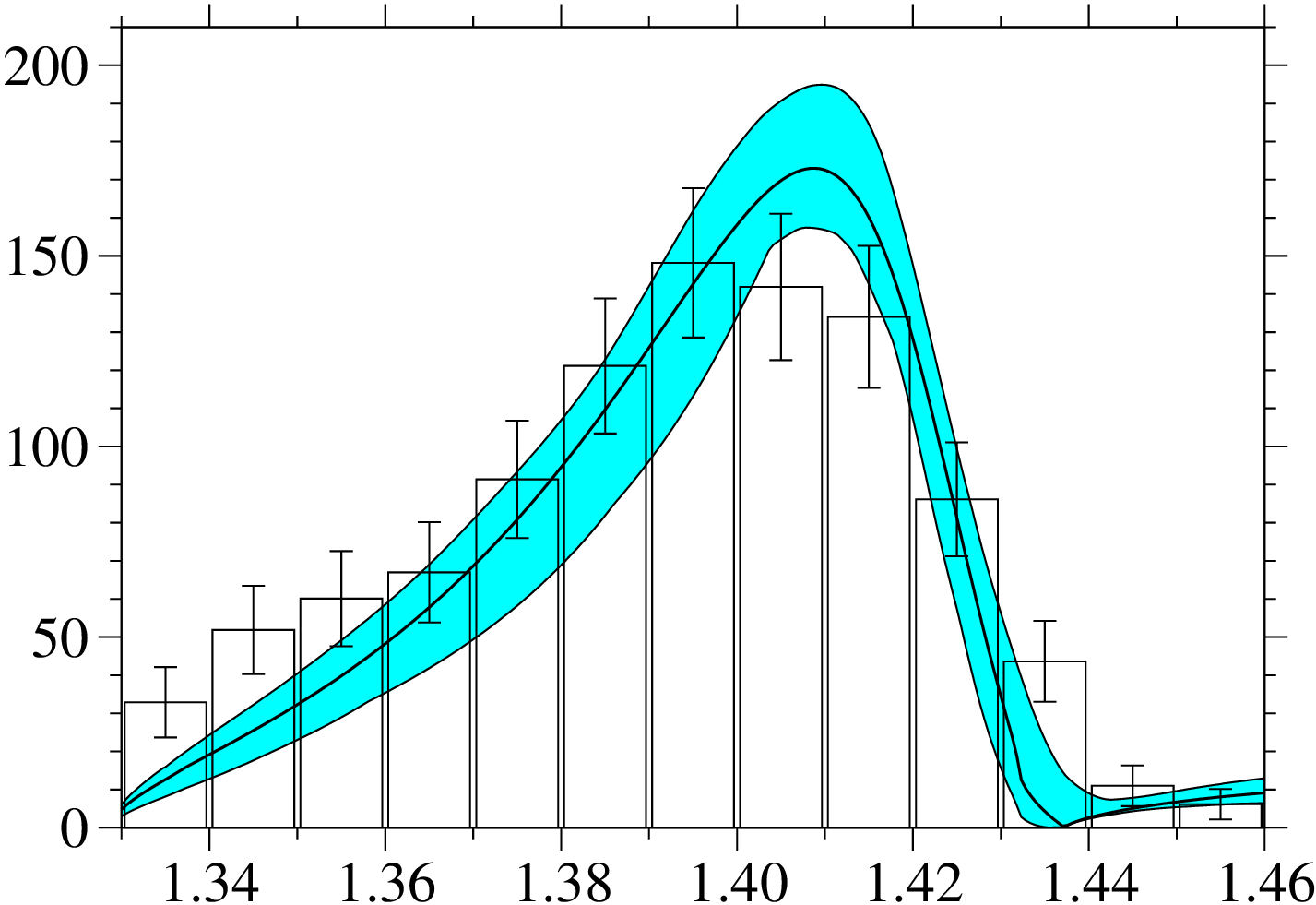}
  \put(-11,19){\rotatebox{90}{{\scalebox{0.9}{arbitrary units}}}}
  \put(40,-7){\scalebox{0.9}{$\sqrt{s}$ \quad (GeV)}}
\end{overpic} \\[5ex]
``WTB'' approach \\[0.5ex]
\begin{overpic}[width=0.40\textwidth,clip]{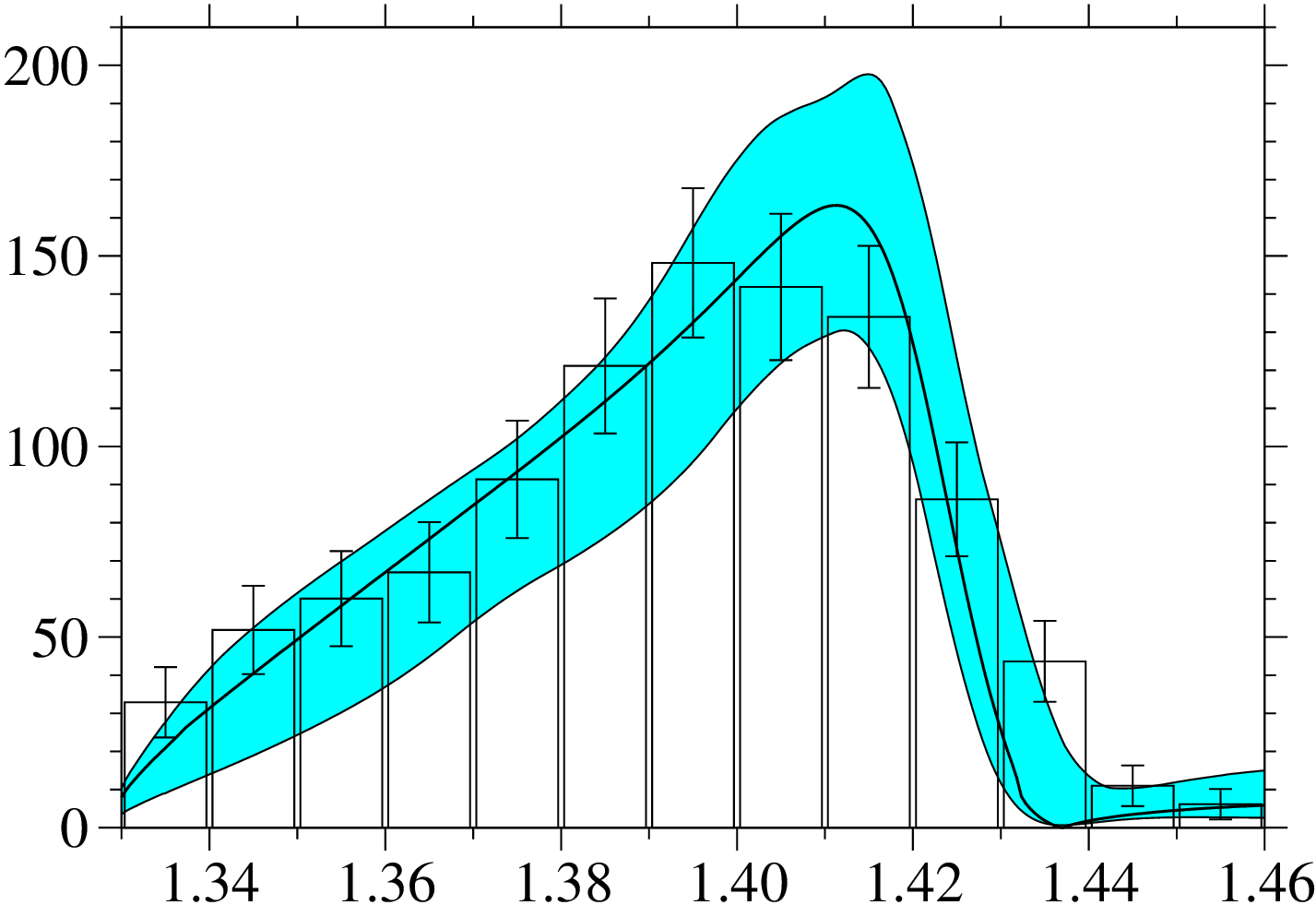}
  \put(-11,19){\rotatebox{90}{{\scalebox{0.9}{arbitrary units}}}}
  \put(40,-7){\scalebox{0.9}{$\sqrt{s}$ \quad (GeV)}}
\end{overpic} \\[5ex]
``full'' approach \\[0.5ex]
\begin{overpic}[width=0.40\textwidth,clip]{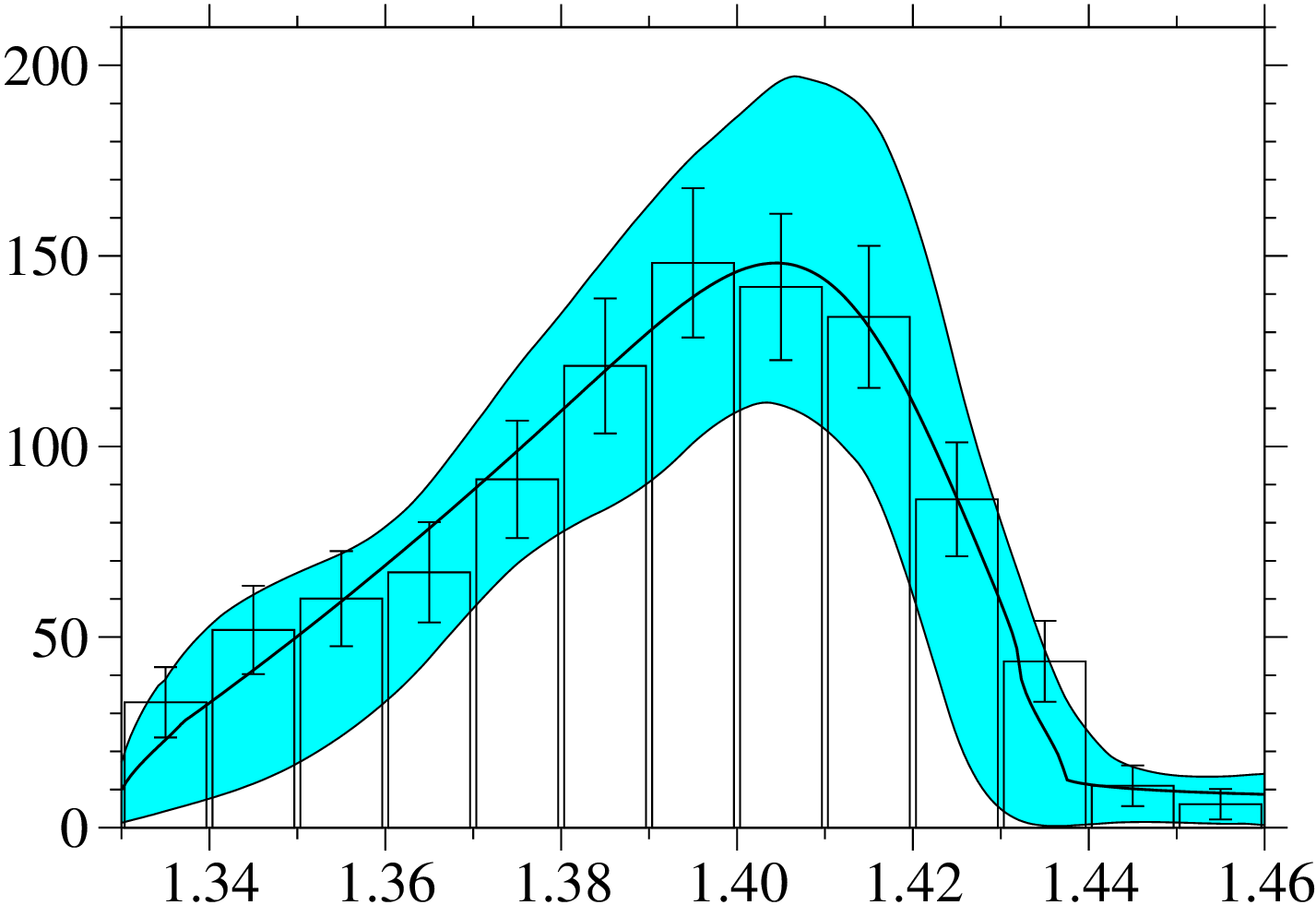}
  \put(-11,19){\rotatebox{90}{{\scalebox{0.9}{arbitrary units}}}}
  \put(40,-7){\scalebox{0.9}{$\sqrt{s}$ \quad (GeV)}}
\end{overpic}
\end{tabular}
\vspace{2.0ex}
\caption{$\pi^- \Sigma^+$ event distribution for the three different approaches.
         The best fit is represented by the solid line, while the shaded
         area indicates the 1$\sigma$ confidence region.  The data are taken from \cite{Hem}
         and supplemented by statistical errors following \cite{DD}.}
\label{fig:PS_OM}
\end{figure}

In the appendix, we present the results of the fits to scattering data. The overall agreement
with the experimental data is very good.
In Figs.~\ref{fig:CSWT}, \ref{fig:CSC}, \ref{fig:CSB} we plot the elastic and inelastic $K^- p$ 
scattering cross sections for the three approaches ``WT'', ``WTB'', and ``full'', respectively, 
while the $\pi \Sigma$ event distribution is shown in Fig.~\ref{fig:PS_OM}.
For the accurately determined threshold branching ratios $\gamma$, $R_c$, $R_n$ defined in
Eq.~\ref{eq:BRdef} we obtain:
\begin{center}
\begin{tabular}{|l||r@{\,}l|r@{\,}l|r@{\,}l||r@{$\,\pm\,$}l|}
\hline
& \multicolumn{2}{c|}{``WT''} & \multicolumn{2}{c|}{``WTB''} & \multicolumn{2}{c||}{``full''}
& \multicolumn{2}{c|}{exp.\ \cite{Now, Tov}} \\
\hline
$\gamma$ & 2.35  & $^{+0.07}_{-0.06}$ 
         & 2.36  & $^{+0.03}_{-0.03}$ 
         & 2.36  & $^{+0.10}_{-0.09}$ 
         & 2.36  & 0.04 \\
\hline
$R_c$    & 0.655 & $^{+0.001}_{-0.018}$ 
         & 0.664 & $^{+0.022}_{-0.024}$
         & 0.663 & $^{+0.016}_{-0.018}$ 
         & 0.664 & 0.011 \\
\hline
$R_n$    & 0.191 & $^{+0.027}_{-0.031}$ 
         & 0.193 & $^{+0.009}_{-0.017}$
         & 0.190 & $^{+0.026}_{-0.036}$ 
         & 0.189 & 0.015 \\
\hline
\end{tabular}
\end{center}
The central values correspond to the best fit in each approach, while the errors indicate the 
$1\sigma$ confidence region.

\end{document}